\documentclass[12pt,english,longbibliography,nofootinbib,superscriptaddress,12pt,sort&compress,showkeys]{revtex4}
\usepackage[latin9]{inputenc}
\usepackage[letterpaper]{geometry}
\usepackage{amsmath}
\usepackage{amssymb}
\usepackage{graphicx}

\makeatletter
\@ifundefined{textcolor}{}
{%
 \definecolor{BLACK}{gray}{0}
 \definecolor{WHITE}{gray}{1}
 \definecolor{RED}{rgb}{1,0,0}
 \definecolor{GREEN}{rgb}{0,1,0}
 \definecolor{BLUE}{rgb}{0,0,1}
 \definecolor{CYAN}{cmyk}{1,0,0,0}
 \definecolor{MAGENTA}{cmyk}{0,1,0,0}
 \definecolor{YELLOW}{cmyk}{0,0,1,0}
}

\usepackage{indentfirst}
\usepackage{amsfonts}
\usepackage[T1]{fontenc}
\usepackage{ae,aecompl}
\usepackage{sidecap}
\usepackage[section]{placeins}
\usepackage{epsf}
\makeatother

\usepackage{babel}
\date{\today}

\usepackage{chngcntr}

\makeatother

\usepackage{babel}
\begin{document}
\begin{center}
\textbf{\large{}Direct atomistic modeling of solute drag by moving
grain boundaries} 
\par\end{center}
\author{\noindent R. K. Koju}
\address{Department of Physics and Astronomy, MSN 3F3, George Mason University,
Fairfax, Virginia 22030, USA}
\author{\noindent Y. Mishin}
\address{Department of Physics and Astronomy, MSN 3F3, George Mason University,
Fairfax, Virginia 22030, USA}
\begin{abstract}
We show that molecular dynamics (MD) simulations are capable of reproducing
the drag of solute segregation atmospheres by moving grain boundaries
(GBs). Although lattice diffusion is frozen out on the MD timescale,
the accelerated GB diffusion provides enough atomic mobility to allow
the segregated atoms to follow the moving GB. This finding opens the
possibility of studying the solute drag effect with atomic precision
using the MD approach. We demonstrate that a moving GB activates diffusion
and alters the short-range order in the lattice regions swept during
its motion. It is also shown that a moving GB drags an atmosphere
of non-equilibrium vacancies, which accelerate diffusion in surrounding
lattice regions.
\end{abstract}
\maketitle

\section{Introduction}

Grain boundaries (GBs) are critical components of microstructure that
often control properties and performance of materials \citep{Balluffi95,Mishin2010a}.
In alloys, GBs can interact with the chemical components by a number
of different mechanisms. A stationary GB can form a segregation atmosphere
that reduces its free energy and alters its mechanical responses,
thermal, electronic, and other properties. When a GB moves under an
applied thermodynamic driving force, its interaction with the alloy
components can drastically change its mobility. 

\begin{figure}[ht]
\noindent \centering{}\includegraphics[width=0.4\textwidth]{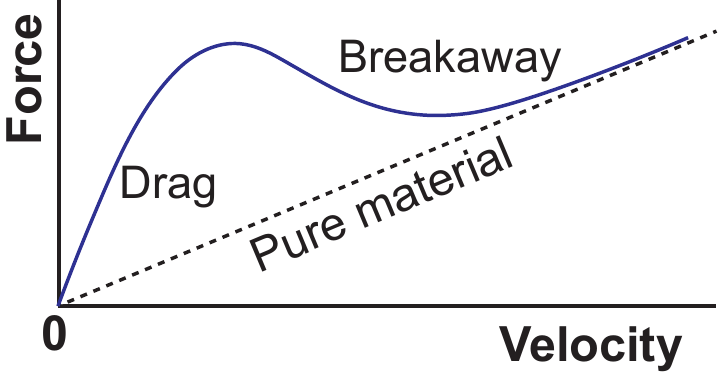}
\caption{Schematic force-velocity diagram according to the classical model
of GB solute drag \citep{Cahn-1962,Lucke-Stuwe-1963,Lucke:1971aa}.
The maximum of the force separates two kinetic regimes. The moving
GB drags the segregation atmosphere at low velocities and breaks away
from it at high velocities. \label{fig:GB17530-1}}
\end{figure}

One of the interaction mechanisms is the drag of the segregation atmosphere
by a moving GB. The effect is known as the GB solute drag and has
been the subject of numerous experimental, theoretical and modeling
studies over the past decades. The first quantitative model of solute
drag was proposed by Cahn \citep{Cahn-1962} and Lücke et al.~\citep{Lucke-Stuwe-1963,Lucke:1971aa}.
Their model predicts a highly nonlinear relation between the GB velocity
and the drag (friction) force, with a maximum force reached at some
critical velocity (Fig.\ \ref{fig:GB17530-1}). On the low-velocity
side of the maximum, the segregation atmosphere moves together with
the GB. Dragging the entire atmosphere requires a larger driving force
than for moving the same GB through a uniform solution. On the high-velocity
side of the maximum, the GB breaks away from the atmosphere and the
drag force drops. Eventually, the GB forms a new, much lighter segregation
atmosphere that poses less resistance to its motion.

The solute drag effect has been studied by a number of simulation
approaches, including the phase field \citep{Wang03,Li:2009aa,Shahandeh:2012aa,Gronhagen:2007aa,Abdeljawad:2017aa}
and phase-field crystal \citep{Greenwood:2012aa} methods, atomic-level
computer simulations \citep{Mendelev02a,Sun:2014aa,Wicaksono:2013aa,Mendelev:2001aa,Rahman:2016aa,Kim:2008aa},
and discrete quasi-1D models \citep{Wang03,Mishin:2019aa}. Another
effective approach to studying the solute drag is offered by molecular
dynamics (MD) simulations. MD provides access to all atomic-level
details of the GB motion and permits continual tracking of both the
GB velocity and the driving force. No approximations are made other
than the atomic interaction model (interatomic potential), which for
some systems can be quantitatively accurate. Unfortunately, the timescale
of MD simulations is presently limited to tens, or at best a hundred,
nanoseconds. This timescale is too short to capture lattice diffusion
by the vacancy mechanism. Given the small concentration of equilibrium
vacancies and the low vacancy-atom exchange rate, an average atom
can only make a few jumps in a typical MD simulation. Meanwhile, the
classical solute drag model \citep{Cahn-1962,Lucke-Stuwe-1963,Lucke:1971aa}
assumes that the solute atoms diffuse toward or away from the moving
GB through the lattice at a finite rate. Accordingly, the lattice
diffusion coefficient is one of the parameters of the model. 

Because of the timescale disparity mentioned above, a common perception
has been that the MD method is not suitable for direct simulations
of the solute drag. More precisely, in MD simulations, it is always
possible to drag a GB through a solid solution. This will generally
require a larger driving force than in a single-component system.
As such, the simulations will capture some solute resistance. However,
this kinetic regime will lie on the far right side of the classical
force-velocity curve (Fig.\ \ref{fig:GB17530-1}). What has been
perceived as unfeasible is to reproduce the low-velocity branch of
the curve, on which the GB moves together with its segregation atmosphere.
To our knowledge, this kinetic regime has not been implemented in
MD simulations so far.

In this paper, we demonstrate that it is, in fact, possible to reproduce
\emph{both} branches of the force-velocity curve by MD simulations,
including the solute drag regime existing at small velocities. Although
lattice diffusion remains frozen out on the MD timescale, the accelerated
GB diffusion provides sufficient atomic mobility to drag the segregation
atmosphere together with the boundary. A high-angle tilt GB moving
in Cu-Ag solid solutions is chosen as a model system. Special attention
is devoted to proving convincing evidence that the GB does indeed
drag the solute atoms, and that it activates diffusion in lattice
regions traversed during its motion. We also show that a moving GB
carries an atmosphere of non-equilibrium vacancies, which contribute
to the acceleration of diffusion processes in surrounding lattice
regions.

\section{Methodology}

\noindent Atomic interactions in the Cu-Ag system were modeled with
a reliable embedded atom potential \citep{Williams06} reproducing
a wide spectrum of properties of Cu, Ag and Cu-Ag phases. In particular,
the potential predicts the Cu-Ag phase diagram in agreement with experiment.
MD simulations were performed using the Large-scale Atomic/Molecular
Massively Parallel Simulator (LAMMPS) \citep{Plimpton95}.

\begin{figure}[ht]
\noindent \centering{}\includegraphics[width=0.5\textwidth]{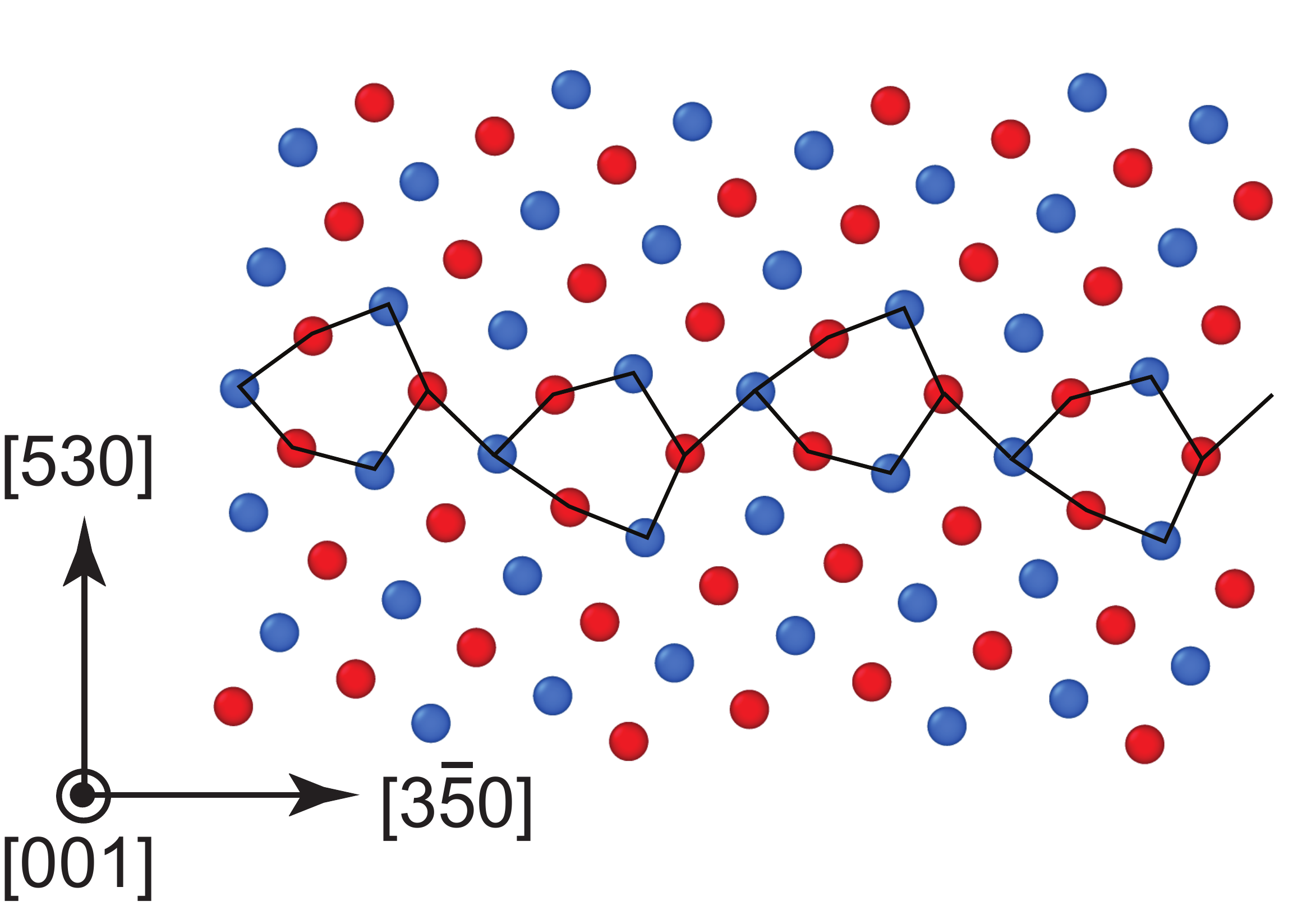}
\caption{Atomic structure of the $\Sigma17(530)[001]$ symmetrical tilt GB
in Cu. The red and blue circles represent the atoms located in alternating
(002) planes normal to the {[}001{]} tilt axis. The structural units
are outlined. \label{fig:GB17530}}
\end{figure}

As a model, we chose the symmetrical tilt $\Sigma17(530)[001]$ GB
with the misorientation angle of $61.93^{\circ}$ ($\Sigma$ is the
reciprocal density of coincident sites, {[}001{]} is the tilt axis,
and $(530)$ is the GB plane). The GB was created in a rectangular
simulation block with edges aligned with the Cartesian axes $x$,
$y$ and $z$, which in turn were normal to the GB tilt axis, and
parallel to the GB tilt axis, and normal to the GB plane, respectively.
The block had the approximate dimensions of $6.32\times6.14\times80.10$
nm$^{3}$ and contained $2.64\times10^{5}$ atoms. The boundary conditions
were initially periodic in all three directions. The ground-state
structure of this GB is known \citep{Suzuki05a,Cahn06b} and consists
of identical kite-shape structural units arranged in a zig-zag array
as shown in Fig.\ \ref{fig:GB17530}. The GB energy is 856 mJ/m$^{2}$.

A uniform solid solution was created by random substitution of Cu
atoms by Ag atoms to achieve the desired alloy composition. The alloy
compositions studied here lay within the Cu-based solid solution domain
on the Cu-Ag phase diagram. Note that the initial state of the solution
did not represent a thermodynamic equilibrium. Under equilibrium conditions,
the GB would create a segregation of Ag atoms as was demonstrated
in previous work \citep{Frolov:2015aa,Frolov:2016aa,Hickman:2016aa,Cu-Ag-segregation-diffusion}.
Furthermore, the solution did not have the correct short range order
that forms upon equilibration. The initially uniform random solution
was chosen here intentionally in order to observe the process of atmosphere
formation during the GB motion. In addition, the formation of short
range order was used as an indicator of solute diffusion activated
by the GB motion. 

The GB was moved by an applied shear strain. The driving force for
the motion arises due to the shear-coupling effect \citep{Cahn04a,Suzuki05b,Cahn06b},
in which relative translations of the grains parallel to the GB plane
cause normal GB motion. Reversely, normal GB motion produces shear
deformation of the lattice region traversed by the motion. The shear-coupling
effect is characterized by the coupling factor $\beta=v_{||}/v$,
where $v_{||}$ is the velocity of parallel translation of the grains
and $v$ is the velocity of normal GB motion. For perfect coupling,
$\beta$ is a constant that only depends on crystallographic characteristics
of the GB. For the particular boundary studied here, the ideal coupling
factor is $\beta=-0.496$ (the negative sign reflects the sign convention
\citep{Cahn06b}). To produce GB motion, the boundary condition in
the $z$-direction was replaced by free surfaces. The simulation block
was re-equilibrated by an MD run in the $NpT$ ensemble (fixed number
of atoms $N$, temperature $T$ and pressure $p=0$). Next, the relative
positions of the atoms within 1.5 nm thick surface layers were fixed.
One bottom layer was fixed permanently, while the other layer was
moved, as a rigid body, with a constant translation velocity $v_{||}$
parallel to the $x$-direction. All other atoms remained dynamic.
The MD ensemble was then switched to $NVT$ (fixed volume $V$) and
the GB was driven in a chosen direction, starting from its initial
position at the center of the block, by applying a given translation
velocity $v_{||}$. 

During the simulations, the GB position was tracked by finding the
peak of potential energy of atoms averaged over thin layers parallel
to the GB plane. Under perfect coupling conditions, the driving force
causing the GB motion equals $P=\beta\tau_{xy}$, where $\tau_{xy}$
is the shear stress parallel to the GB plane \citep{Cahn04a,Cahn06b}.
The latter was computed from the virial equation averaged over the
entire simulation block and over time in the steady-state regime.
The redistribution of Ag atoms caused by the GB motion was analyzed
in detail using the OVITO visualization software \citep{Stukowski2010a}.
The simulations were performed at the temperature of 1000 K and alloy
compositions varying from pure Cu to the solidus line. The grain translation
velocity was varied between 0.01 and 60 m/s. Implementing lower velocities
would require computational resources that were not available to us.

\section{Results}

\subsection{The solute drag dynamics}

\begin{figure}[ht]
\noindent \centering{}\includegraphics[width=0.6\textwidth]{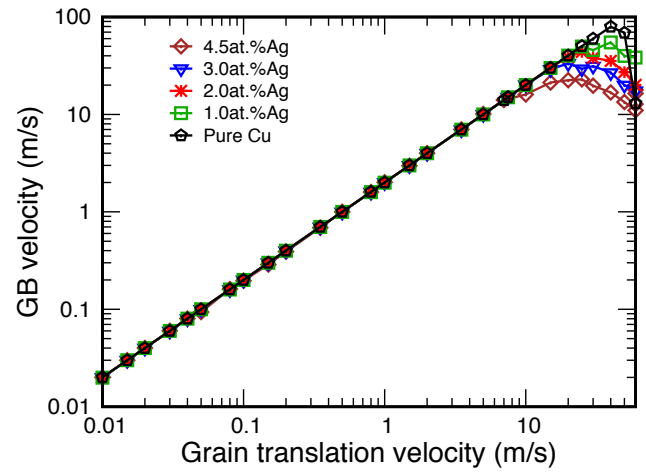}
\caption{GB migration velocity $v$ as a function of the applied grain translation
velocity $v_{||}$. The alloy compositions are indicated in the key.\label{fig:coupling factor}}
\end{figure}

The ability to predict the driving force using the equation $P=\beta\tau_{xy}$
predicates on the shear coupling being perfect. Our first step was,
therefore, to find the conditions under which the coupling factor
$\beta$ was close to its ideal value. Fig.\ \ref{fig:coupling factor}
shows a typical plot of GB velocity versus the imposed grain translation
velocity. At small enough velocities, the relation between $v$ and
$v_{||}$ is linear with the slope matching the ideal value of $\beta$.
After a critical velocity is reached, the GB shows downward deviations
from the ideal behavior. Such deviations were observed previously
\citep{Cahn06b,Mishin07a} and are usually caused by occasional switches
between coupled motion and sliding events. Once the boundary enters
this regime, the driving force can no longer be extracted from the
shear stress. We thus limited the present simulations to GB velocities
below the onset of the downward deviations. For pure Cu, the highest
velocity studied was about 40 m/s. For the alloys, the upper bound
was lower (10 to 20 m/s), depending on the alloy composition.

\begin{figure}[ht]
\noindent \centering{}\includegraphics[width=0.6\textwidth]{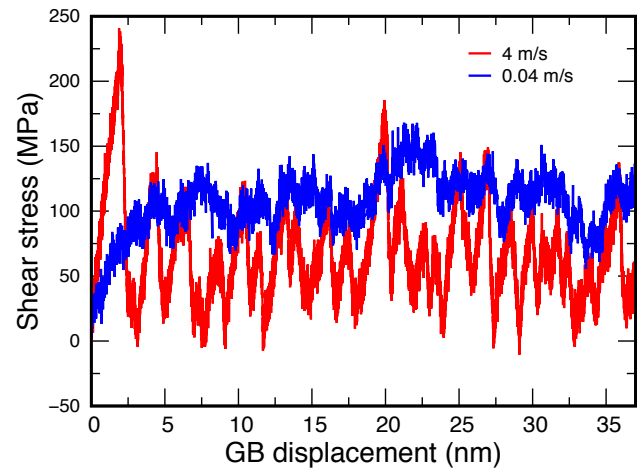}
\caption{Representative stress-time curves in the Cu-2at.\%Ag alloy at two
different GB velocities. \label{fig:stress-time}}
\end{figure}

Upon application of a grain translation velocity, the shear stress
$\tau_{xy}$ initially showed a transient behavior followed by a steady
state regime in which $\tau_{xy}$ continued to fluctuate around a
constant average value (Fig.~\ref{fig:stress-time}). The transient
time period was somewhat longer for the alloys than for pure Cu. Since
the solid solution was initially uniform, time was required to form
a dynamic segregation atmosphere surrounding the moving boundary.
The shear stresses, and thus the driving forces reported below, were
obtained by averaging over a period of time after the onset of the
steady state regime.

\begin{figure}[ht]
\noindent \begin{centering}
(a) \includegraphics[width=0.6\textwidth]{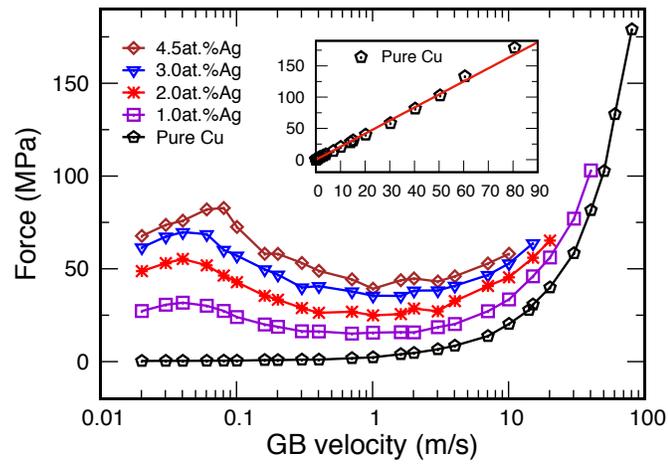} 
\par\end{centering}
\noindent \begin{centering}
\bigskip{}
\par\end{centering}
\noindent \begin{centering}
(b)\includegraphics[width=0.6\textwidth]{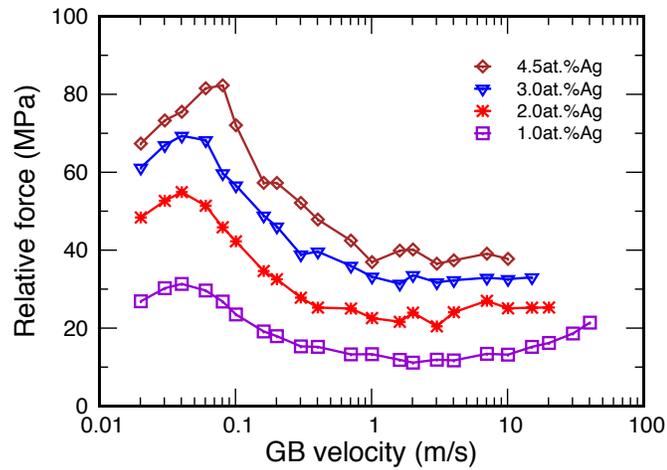}
\par\end{centering}
\caption{(a) Driving force for GB motion as a function of GB velocity at different
alloy compositions. The inset is a zoom into the linear part of the
force-velocity relation in pure Cu. (b) The same functions but the
driving force in pure Cu has been subtracted from the driving force
in the alloys. \label{fig:solutedrag}}
\end{figure}

The central result of this work is presented in Fig.~\ref{fig:solutedrag},
where the driving force for GB motion is plotted against the GB velocity.
Since the GB velocities span three orders of magnitude, they are shown
on the logarithmic scale. In pure Cu, the initially linear force-velocity
relation becomes highly nonlinear at high velocities. Even though
the GB remains perfectly coupled and moves by the same atomic mechanism,
the resistance to its motion sharply increases at high velocities.
This non-linearity has been seen previously and is caused by a combination
of factors, such as (1) faster-than-linear suppression of the energy
barrier as a function of the force \citep{Ivanov08a}, and (2) finite
rate of heat dissipation by the moving boundary (phonon drag). This
nonlinear regime was not considered in the classical model \citep{Cahn-1962,Lucke-Stuwe-1963,Lucke:1971aa}.
In the alloys, the driving force also accelerates in the high velocity
limit. However, the force-velocity relation in the alloys is different
from that in Cu in two major ways:
\begin{itemize}
\item The driving force in alloys is systematically higher than that in
pure Cu and increases with the solute concentration. In other words,
the driving force required for moving the GB with a given velocity
increases with the solute concentration. This solute resistance effect
is demonstrated more clearly in Fig.~\ref{fig:solutedrag}b, where
the driving force existing in pure Cu has been subtracted from that
in the alloys.
\item The solute resistance exhibits a maximum at some velocity that increases
with the solute concentration. This is precisely the behavior predicted
by the classical solute drag models \citep{Cahn-1962,Lucke-Stuwe-1963,Lucke:1971aa}
(cf.~Fig.~\ref{fig:GB17530-1}). Thus, on the left of the maximum,
the GB is expected to drag of the segregation atmosphere with it,
while on the right of the maximum its breaks away from the atmosphere.
\end{itemize}
The curves presented in Fig.~\ref{fig:solutedrag} clearly demonstrate
that we have been able to reproduce by MD simulations the two modes
of the solute resistance to GB motion: both the breakaway mode at
high velocities and the solute drag mode at low velocities.

\subsection{Direct evidence of solute drag}

The fact that the present simulations capture the drag of the solute
atoms by the GB can be demonstrated directly by tracking the positions
of the solute atoms. One such test is illustrated in Fig.~\ref{fig:GBdrag}
for the Cu-2at.\%Ag alloy. Ag atoms (colored in magenta) located in
two parallel stripes on either side of the initial GB position were
selected for tracking. The GB was driven to the left, eventually passing
through the left-hand stripe. The right-hand stripe was not affected
by the GB motion and only served as a control region. It was found
that the atomic positions in the control region did not practically
change, confirming the absence of lattice diffusion on the MD timescale.
(To be precise, some atoms did make an occasional jump caused by non-equilibrium
vacancies that will be discussed later.) By contrast, the atoms overrun
by a slowly moving boundary were randomly scattered due to the accelerated
GB diffusion. Importantly, the scattering occurs predominantly in
the direction of GB motion, demonstrating that the GB binds the Ag
atoms and drags them along over some distance. In other words, the
selected Ag atoms are drawn into the dynamics segregation atmosphere
traveling together with the boundary. Eventually, they drop out of
the atmosphere due to the random diffusive jumps.

The drag effect can be quantified by plotting the final distribution
of the Ag atoms in the left-hand stripe after it has been traversed
by the boundary as a function of GB velocity and alloy composition
(Fig.~\ref{fig:GBdragB}). Each distribution function was obtained
by averaging over 20-25 passes of the GB through the stripe. The scatter
of the points reflects the limited statistics due to the low solute
concentration. The plots demonstrate that, as the GB velocity decreases,
the concentration profile becomes increasingly asymmetric and develops
a long tail in the low-velocity limit. At the smallest velocity (0.02
m/s), the tail is a factor of 5 longer than the initial width of the
stripe. The effect of the alloy composition on the concentration profiles
is less significant, which is not surprising given that the alloys
are dilute. 

For comparison, Fig.~\ref{fig:GBdragB-1} shows similar concentration
profiles for the host Cu atoms initially residing within the same
stripe. At low GB velocities, the Cu profiles broaden and develop
an asymmetric $\lambda$-shape. However, the broadening is much smaller
than for the Ag atoms and does not contain the long tail in the direction
of motion. The degree of broadening does not practically depend on
the alloy composition. As shown in the Appendix, the characteristic
$\lambda$-shape of the Cu concentration profiles is consistent with
purely diffusional broadening caused by accelerated diffusion during
the time when the GB is inside the stripe. This type of broadening
is not related to solute drag. For Cu atoms, the drag does not exist
in pure Cu and is negligible in alloys, where the Cu segregation (to
be precise, ``anti-segregation'', because the segregated Ag atoms
substitute for Cu) is small. The drastic difference between the profile
of Ag and Cu (cf.~Figs.~\ref{fig:GBdragB} and \ref{fig:GBdragB-1})
confirms that the Ag atoms were subject to a strong drag effect.

\begin{figure}[ht]
\noindent \begin{centering}
(a) \includegraphics[width=0.6\textwidth]{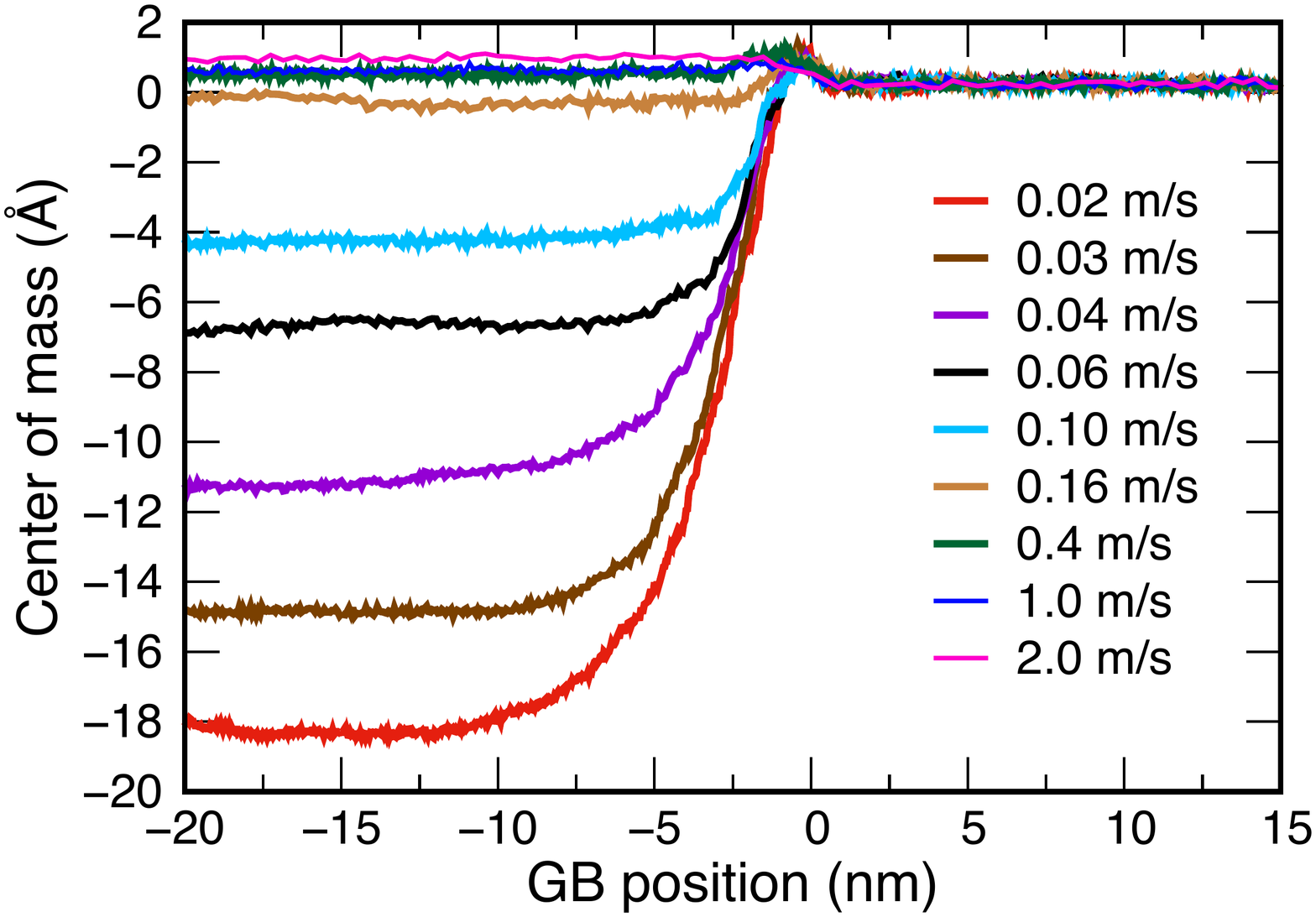}
\par\end{centering}
\noindent \begin{centering}
\bigskip{}
\par\end{centering}
\noindent \centering{}(b) \includegraphics[width=0.6\textwidth]{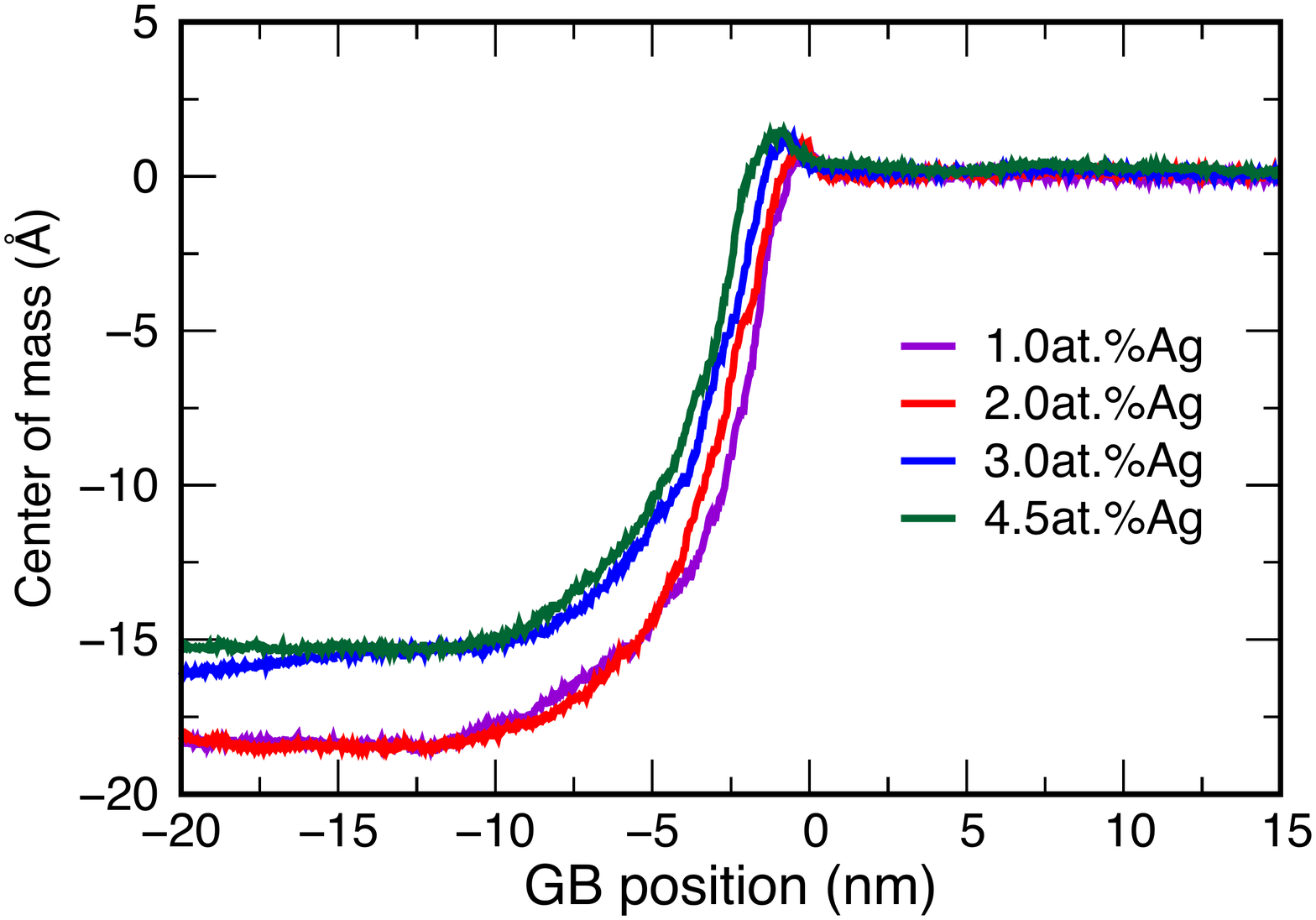}
\caption{Position of the center of mass of Ag atoms located within a stripe
overrun by a moving GB as a function of GB position $z$. The GB moves
from right to left with the center of the stripe located at $z=0$.
(a) for different GB velocities at a fixed alloy composition (Cu-2at.\%Ag);
(b) for different alloy compositions at a fixed velocity of 0.02 m/s.\label{fig:CM-Ag}}
\end{figure}

Yet another demonstration of the solute drag was obtained by tracking
the $z$-coordinate of the center of mass of the selected atoms. By
definition of the solute drag, the GB carries over some distance the
center of mass of any group of solute atoms that it overrun during
its motion. This was indeed observed at small GB velocities as illustrated
in Fig.~\ref{fig:CM-Ag}. The center of mass of the Ag atoms located
within the selected stripe initially coincides with the center of
the stripe ($z=0$). As soon as the moving GB enters the stripe, the
center of mass of the Ag atoms starts moving in the same direction.
This motion persists for some time after the GB exists the stripe,
demonstrating that the Ag atoms continued to travel along with the
boundary. Since they also make random diffusive jumps, they gradually
drop out the segregation atmosphere and the center of mass eventually
stops moving. At slow GB velocities, the net displacement of the center
of mass reaches 2 nm, which is at least twice the GB width.

\begin{figure}[ht]
\noindent \begin{centering}
(a) \includegraphics[width=0.6\textwidth]{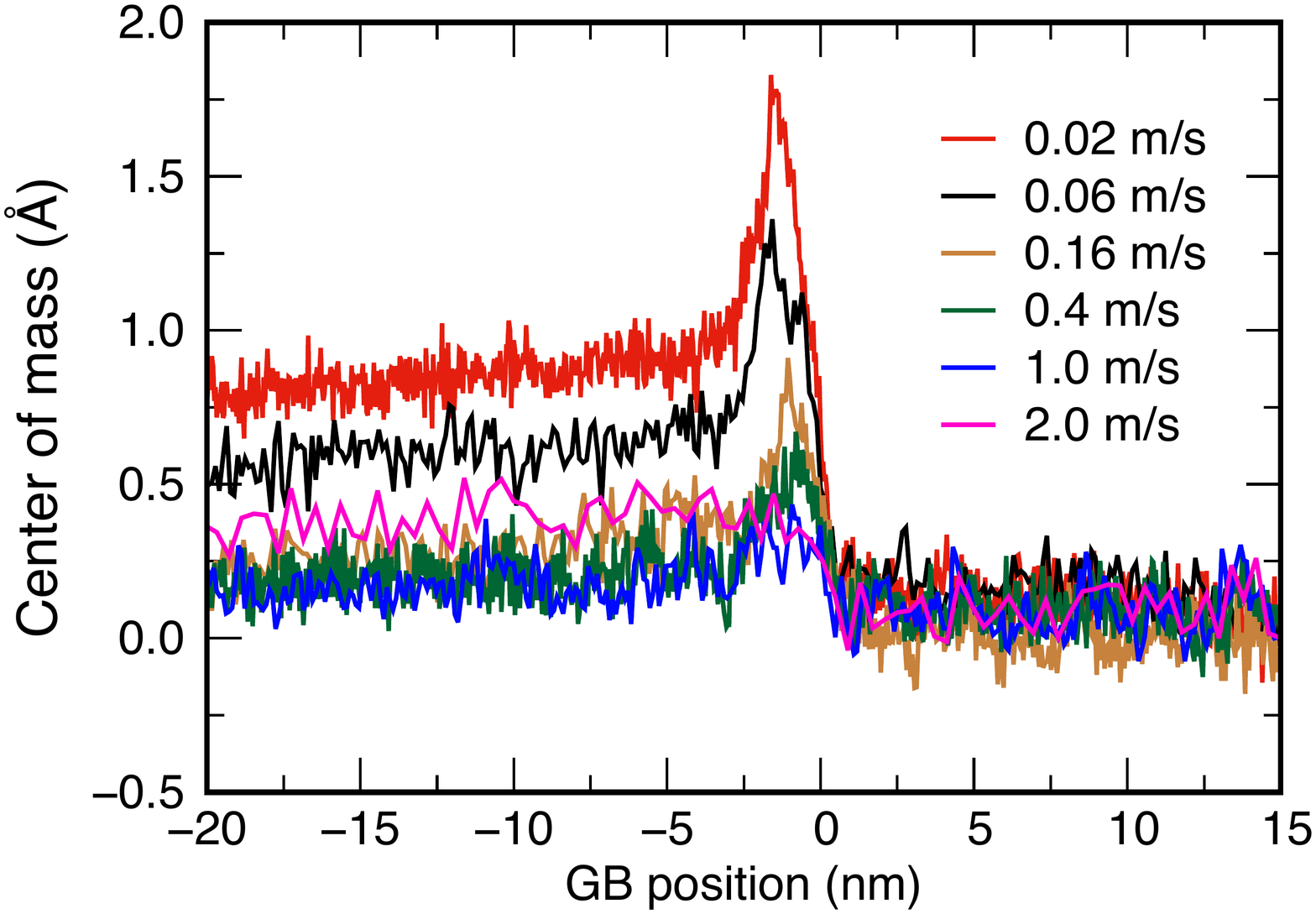}
\par\end{centering}
\noindent \begin{centering}
\bigskip{}
\par\end{centering}
\noindent \centering{}(b) \includegraphics[width=0.6\textwidth]{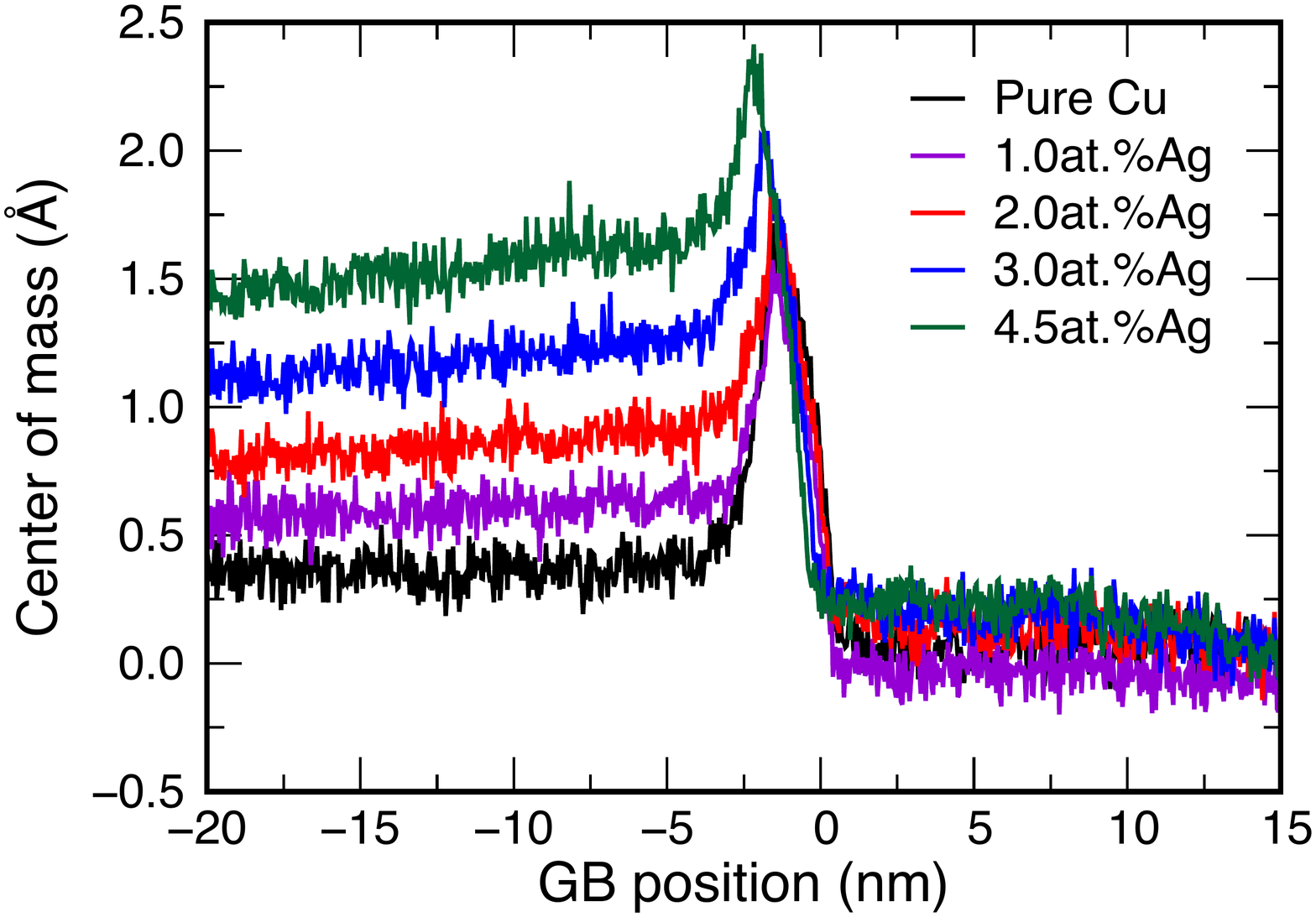}
\caption{Position of the center of mass of Cu atoms located within a stripe
overrun by a moving GB as a function of GB position $z$. The GB moves
from right to left with the center of the stripe located at $z=0$.
(a) for different GB velocities at a fixed alloy composition (Cu-2at.\%Ag);
(b) for different alloy compositions at a fixed velocity of 0.02 m/s.\label{fig:CM-Cu}}
\end{figure}

In comparison, the net displacement of the center of mass of the Cu
atoms located in the same stripe is drastically smaller (Fig.~\ref{fig:CM-Cu}).
Even at the lowest GB velocity, the displacement of the center of
mass is on the order of 0.1 nm. Note also that the plots in Figs.~\ref{fig:CM-Ag}
and \ref{fig:CM-Cu} display a characteristic local maximum arising
when the GB is inside the stripe. This small peak is a manifestation
of a subtle diffusion effect that is not related to solute drag. Purely
diffusion calculations presented in the Appendix give a similar small
displacement of the center of mass, followed by its reversal after
the GB exists the stripe. The predicted split of the maximum in two
could not be resolved in the present simulations due to statistical
errors. 

\subsection{The role of grain boundary diffusion}

\begin{figure}[ht]
\noindent \begin{centering}
(a) \includegraphics[width=0.8\textwidth]{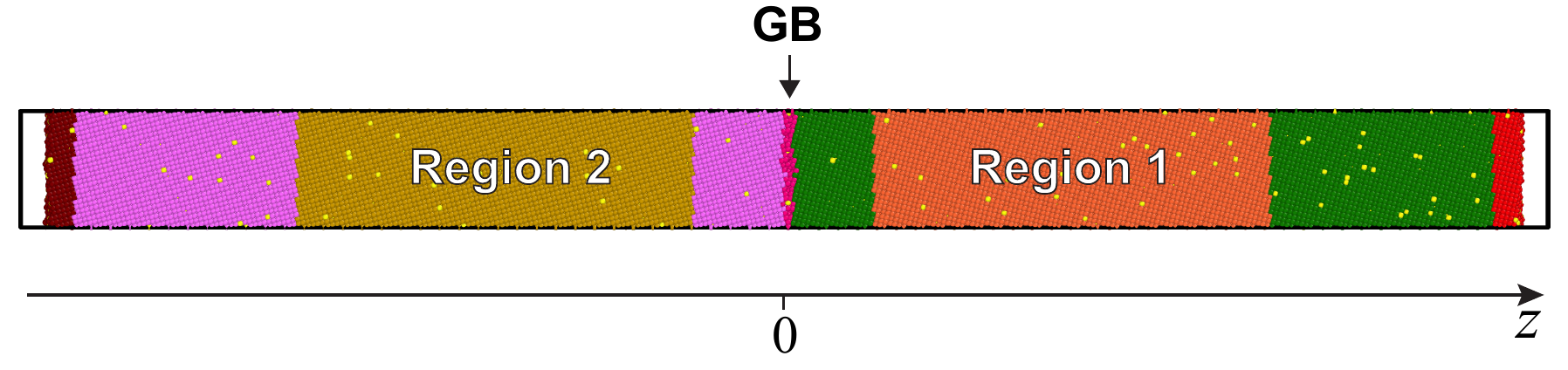}
\par\end{centering}
\noindent \begin{centering}
\bigskip{}
\par\end{centering}
\noindent \begin{centering}
(b) \includegraphics[width=0.6\textwidth]{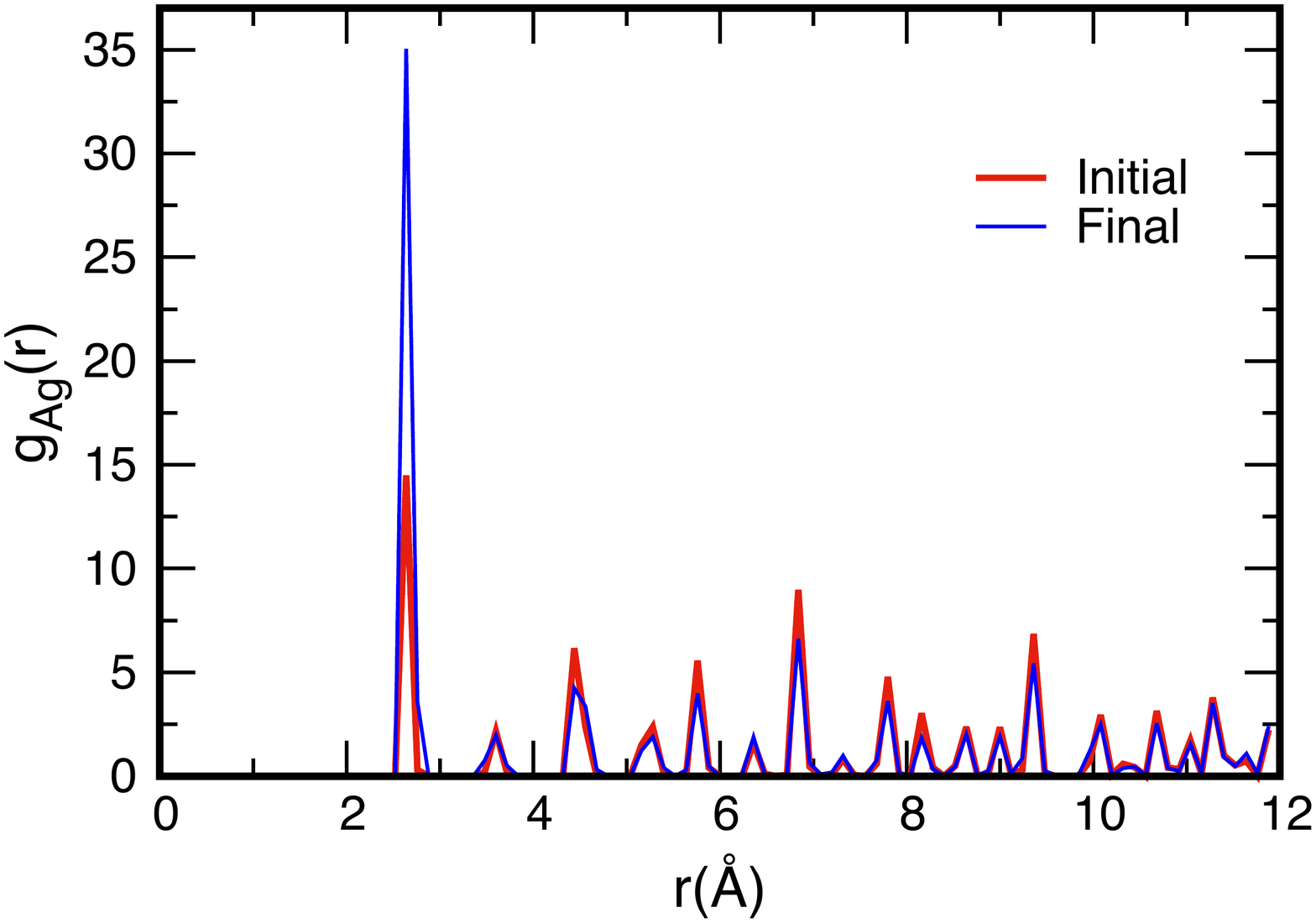} 
\par\end{centering}
\noindent \begin{centering}
\bigskip{}
\par\end{centering}
\noindent \centering{}(c) \includegraphics[width=0.6\textwidth]{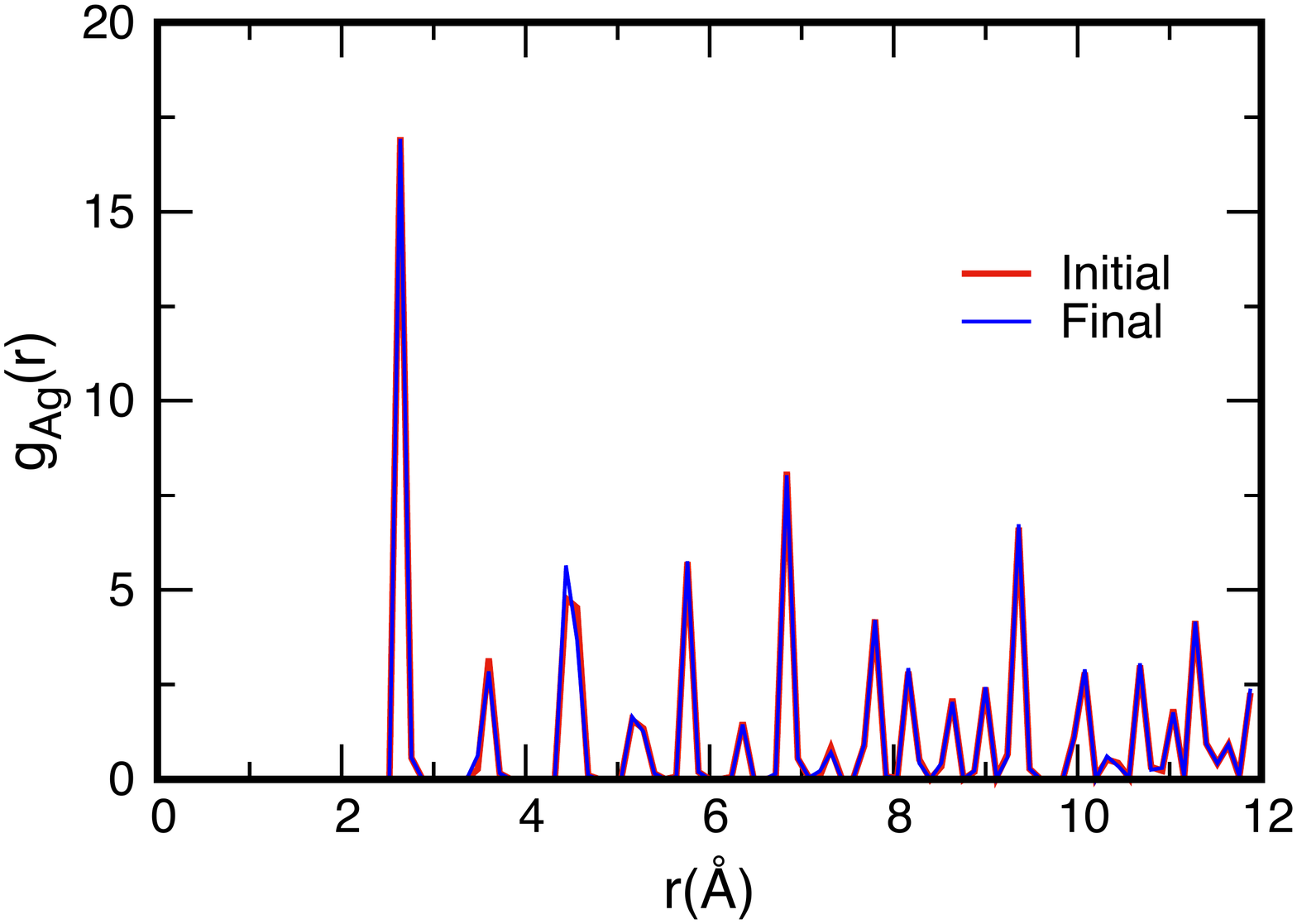}\caption{(a) Selection of regions 1 and 2 in the simulation block. Region 1
is traversed by the GB moving to the right starting from the middle
of the simulation block, while region 2 is a control region that is
not influenced by the GB. (b) and (c) show the final radial distribution
functions of Ag atoms in regions 1 and 2, respectively. The alloy
composition is Cu-2at.\%Ag. \label{fig:RDF}}
\end{figure}

We next demonstrate that a moving GB accelerates diffusion in lattice
regions traversed during its motion. As an indicator of lattice diffusion,
we choose the change in the short range order in the solid solution.
Recall that the Cu-Ag solution was created by random atomic substitution.
As such, it initially has zero short range order. Since lattice diffusion
is frozen, short range order cannot form. To demonstrate the GB effect
on short range order, two regions were selected on either side of
the GB (Fig.~\ref{fig:RDF}a). The GB starts out in the middle of
the simulation block and moves to the right, passing through region
1. Region 2 is not influenced by the GB motion and only serves as
a control region. The radial distribution function (RDF) of Ag atoms
is monitored in both regions as a function of time.

The initial RDF in both regions is similar to that of a single-component
face centered cubic lattice (Fig.~\ref{fig:RDF}b,c). The relative
positions and heights of the peaks are only dictated by the crystallography.
The absolute positions and heights may slightly vary with temperature
due to thermal expansion and with chemical composition due to the
atomic size difference between Cu and Ag. However, in this work such
variations are immaterial because the Ag concentrations are too small
and the temperature is fixed. As the GB sweeps through region 1, the
peak heights change significantly. In particular, the first peak grows
higher, indicating the preference for Ag atoms to become nearest neighbors
(Fig.~\ref{fig:RDF}b). This clustering trend is consistent with
the existence of a wide miscibility gap on the Cu-Ag phase diagram.
Meanwhile, no change in the RDF was found in the control region.

\begin{figure}[ht]
\noindent \begin{centering}
(a) \includegraphics[width=0.6\textwidth]{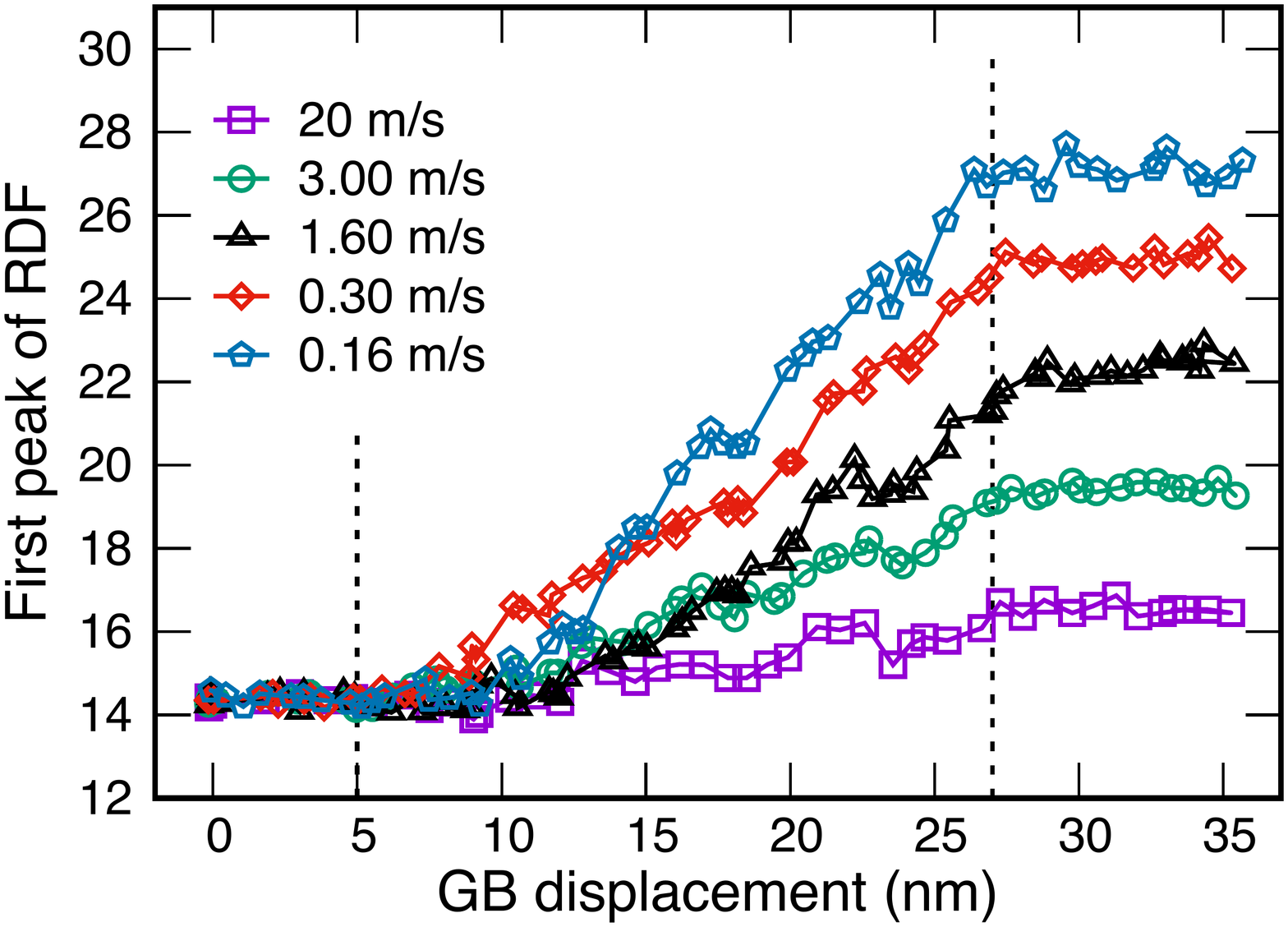} 
\par\end{centering}
\noindent \begin{centering}
\bigskip{}
\bigskip{}
\par\end{centering}
\noindent \centering{}(b) \includegraphics[width=0.63\textwidth]{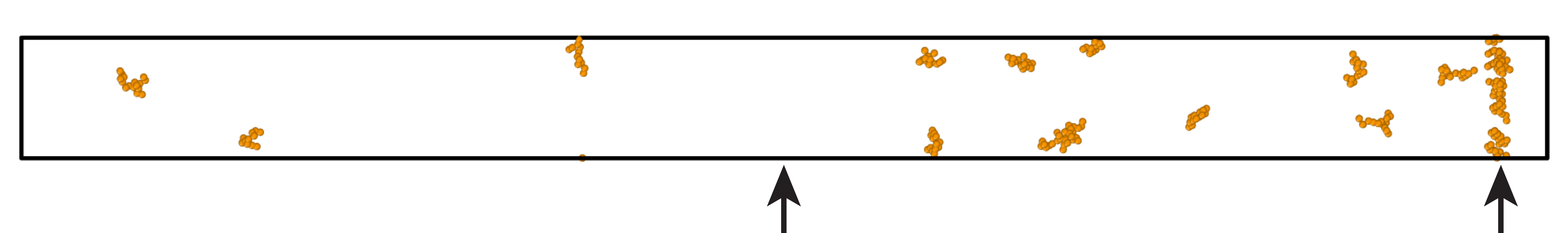}\caption{(a) The height of the first peak of the RDF of Ag atoms in the region
traversed by the moving GB. The dashed vertical lines outline the
boundaries of the region. The alloy composition is Cu-2at.\%Ag. (b)
Formation of Ag clusters in the region swept by the GB moving in the
Cu-4.5at.\%Ag alloy with the velocity of 0.04 m/s. Only Ag clusters
containing 15 or more atoms are shown, while all other atoms are invisible.
The arrows indicate the initial (left) and final (right) GB positions.
\label{fig:Peak-evolution}}
\end{figure}

In Fig.~\ref{fig:Peak-evolution}a, the height of the first peak
of the RDF averaged over region 1 is plotted as a function of GB position.
The plot shows that, as soon as the GB enters the region, the height
of the peak begins to rise. The rise continues until the boundary
exits the region, after which the height of the peak remains constant.
The net change of the peak is small when the boundary moves fast but
increases significantly as the GB velocity decreases. For a visual
demonstration of the short range order formation, Fig.~\ref{fig:Peak-evolution}b
shows the distribution of Ag clusters in the alloy. A cluster is defined
as a group of atoms interconnected by nearest-neighbor bonds. The
probability of such clusters in the initial random alloy is nonzero
but very small. The image demonstrates the formation of a trail of
new clusters containing more than 15 atoms each that were left behind
the moving GB. The formation of short range order in region 1 could
only occur by diffusive jumps of both Ag and Cu atoms. Thus, the simulations
confirm that a GB can activate otherwise frozen diffusion in lattice
regions swept by its motion. The slower the motion, the longer is
the time spent by atoms in the high-diffusivity GB region, and thus
the higher is the effective, GB-induced diffusivity in the lattice.

\subsection{Vacancy generation by a moving grain boundary}

\begin{figure}[h!]
\noindent \begin{centering}
(a)\includegraphics[width=0.47\textwidth]{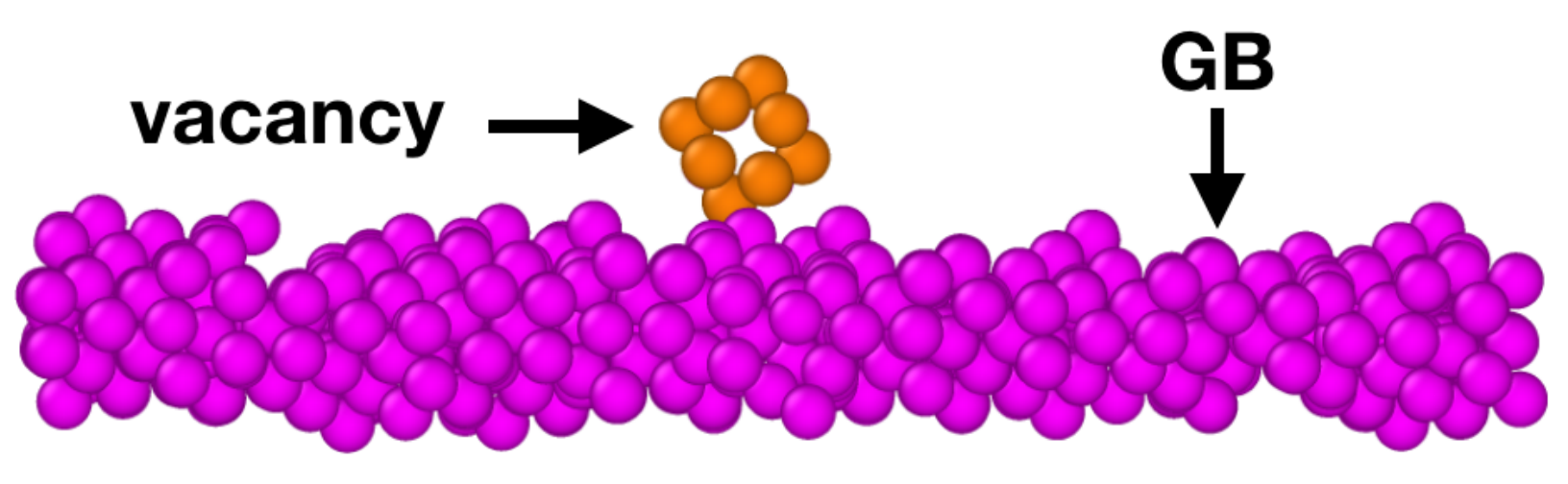} 
\par\end{centering}
\bigskip{}
\bigskip{}

\noindent \begin{centering}
(b) \includegraphics[width=0.7\textwidth]{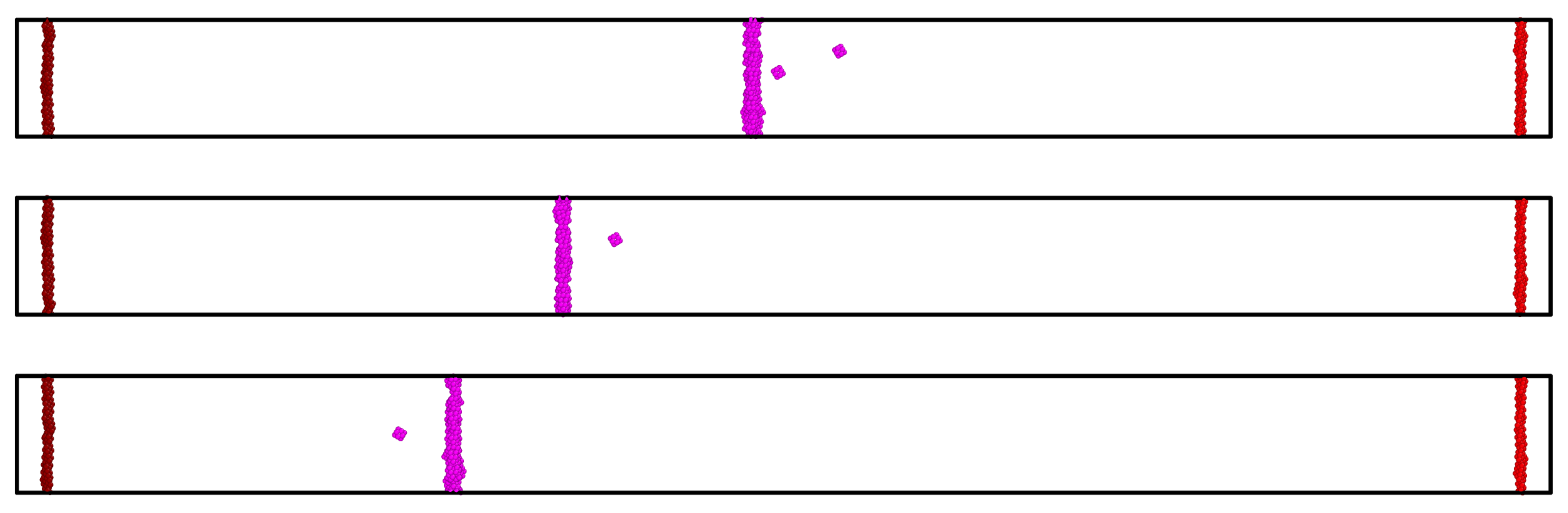}
\par\end{centering}
\noindent \centering{}\caption{Example of vacancy ejection and absorption during GB motion. (a) The
GB moves in the downward direction and the vacancy is at the center
of the 12-atom cluster of defected atoms revealed by common neighbor
analysis with OVITO \citep{Stukowski2010a}. The GB orientation is
the same as in Fig.~\ref{fig:GB17530} (b) Sequence of snapshots
of a GB moving to the left with the velocity of 0.02 m/s. The dots
represent the non-equilibrium vacancies surrounding the GB.\label{fig:GBvacancy}}
\end{figure}

The simulations have also revealed that a moving GB spontaneously
emits and re-absorbs vacancies, effectively creating a vacancy atmosphere
moving together with the boundary (Fig.~\ref{fig:GBvacancy}). The
local vacancy concentration in the atmosphere exceeds the equilibrium
vacancy concentration in the material. 

\begin{figure}[h!]
\noindent \begin{centering}
(a)\includegraphics[width=0.5\textwidth]{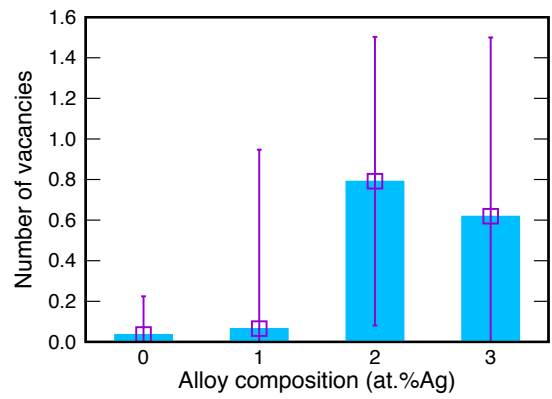}
\par\end{centering}
\bigskip{}

\noindent \begin{centering}
(b) \includegraphics[width=0.6\textwidth]{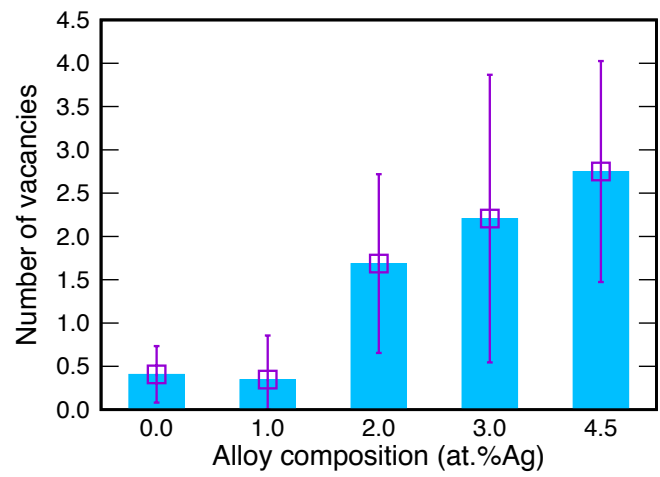}
\par\end{centering}
\noindent \centering{}\caption{Time-averaged number of vacancies in the simulation block containing
(a) stationary GB and (b) moving GB in pure Cu and in Cu-Ag alloys.
The GB velocity is 0.02 m/s. The error bars indicate one standard
deviation. \label{fig:GBvacancy-1}}
\end{figure}

The equilibrium number of vacancies in the simulation block can be
estimated from the equation
\begin{equation}
N_{v}=N\exp\left(-\dfrac{E_{v}}{k_{B}T}\right),\label{eq:N_v}
\end{equation}
where $N$ is the number of atoms in the system, $E_{v}=1.272$ eV
is the vacancy formation energy in Cu predicted by the interatomic
potential, $k_{B}$ is Boltzmann's constant, and $T=1000$ K is temperature.
From this equation, $N_{v}\approx0.1$, meaning that the chance of
seeing a single vacancy in the simulation block is about 10\%. To
verify this estimate, MD simulations were performed on a stationary
GB in both pure Cu and the alloys at 1000 K. In this setting, the
GB serves as a source of vacancies. The time-averaged number of vacancies
present in system is shown in Fig.~\ref{fig:GBvacancy-1}a for several
chemical compositions. The error bar is large because the system typically
contains either one or no vacancy at any given time. In pure Cu, the
average number of vacancies is found to be less than 0.1. This number
increases in the alloys, reaching about 0.7 at 2-3 at.\%Ag.

Fig.~\ref{fig:GBvacancy-1}b shows that a moving GB increases the
number of vacancies in both Cu and the alloys, reaching 2 to 3 vacancies
at 2 to 4.5 at.\%Ag. Visualization of the vacancies reveals that this
increase primarily comes from vacancies ejected by the GB into the
lattice and returning back to it after a short excursion. Thus, the
averaging of the number of vacancies over the entire system strongly
underestimates the actual increase in the vacancy concentration near
the GB. 

\begin{figure}[h!]
\noindent \centering{}\includegraphics[width=0.6\textwidth]{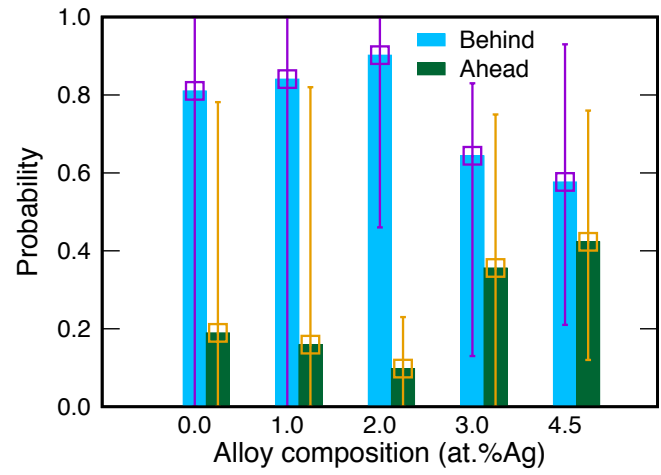}
\caption{Probability of finding a vacancy ahead or behind the moving GB in
the Cu-2at.\%Ag alloy. The GB velocity is 0.02 m/s. The error bars
indicate one standard deviation.\label{fig:vacancyB}}
\end{figure}

It was also noticed that the vacancy distribution on either side of
the moving boundary was not equal, with more vacancies trailing the
GB than jumping ahead of it. This effect is illustrated in Fig.~\ref{fig:vacancyB}
where the probability of finding a vacancy ahead of the moving GB
versus in it wake is shown for several alloy compositions. The asymmetry
is very significant but tends to decrease with the solute concentration.

\section{Concluding remarks}

The main conclusion of this work is that it is presently possible
to model the solute drag by moving GBs by means of conventional MD
simulations. Both the drag and the breakaway kinetic regimes can be
reproduced and studied in atomic detail as a function of alloy composition.
The two regimes are separated by a maximum of the drag force (Fig.~\ref{fig:solutedrag})
in full agreement with predictions of the classical solute drag model
\citep{Cahn-1962,Lucke-Stuwe-1963,Lucke:1971aa}. This finding opens
a research direction in which the solute drag can be investigated
with atomic precision without any approximations other than those
of the interatomic potential. 

Further progress in this direction depends on the availability of
reliable interatomic potentials capable of reproducing thermodynamic
properties of alloys systems on a quantitative level. Another limiting
factor is the computational power. The smallest GB velocity implemented
in this work (0.02 m/s) could have been lower if we had access to
more advanced computational resources, which are certainly available
today. As computers become faster and/or longer MD times become more
accessible, it should soon be possible to extend the solute-drag branch
of the force-velocity curve (Fig.~\ref{fig:solutedrag}) to lower
velocities. It is in that region that some of the most interesting
processes can unfold, such as the drag of a heavy, nearly equilibrium,
segregation atmosphere by a slowly moving boundary. Furthermore, the
cross-section of the GB could be increased to observe the morphological
evolution of the moving boundary. 

The reason why it is possible to observe the solute drag in spite
of the fact that the lattice diffusion is frozen out on the MD timescale
is that the accelerated GB diffusion provides enough atomic mobility
to allow the solute atoms to follow the moving boundary. GB diffusion
is known to be many orders of magnitude faster than lattice diffusion
\citep{Kaur95,Mishin2010a,Mishin:2015ac}. A moving GB effectively
activates solute diffusion in lattice regions swept during its motion.
In this work, it was not possible to quantify this effect directly
due to the small solute concentration. However, indirect evidence
has been provided by showing that the moving GB creates short range
order in the initially random solid solution. Since short range order
formation is a diffusion process, it could only be created by activation
of diffusive jumps. Under real experimental conditions, lattice diffusion
can also contribute to the solute transport toward or away from the
moving GB, at least for slow GB motion. This does not change the solute
drag process qualitatively but may shift the maximum of the drag force
toward higher velocities.

It should be noted that the short range order formation behind a moving
GB gives rise to an additional driving force for forward motion. This
force is equal to the difference between the free energies of the
ordered and disordered states of the solution per unit volume. Attempts
were made to evaluate this force by interrupting the GB motion (removing
the applied shear stress) and continuing the MD simulation with a
stationary GB separating the ordered and disordered lattices. No significant
GB motion was observed, indicating the driving force of ordering was
too small detect it in this work. But it could, in principle, be one
of the factors in microstructure evolution in alloys.

Another effect found in this work is the vacancy generation by a moving
GB (Fig.~\ref{fig:GBvacancy}). The accepted paradigm is that the
fast GB diffusion is localized within the GB core whose width is around
0.5 to 1 nm \citep{Kaur95}. Hence, lattice atoms can only experience
accelerated diffusion when they find themselves inside the core region
of the moving boundary. We find that this picture is incomplete. A
moving GB drags an atmosphere of non-equilibrium vacancies that enhance
the diffusivity in wider lattice regions than the width of the GB
core. Spontaneous ejection of vacancies into the lattice and their
return to the GB after a short excursion appears to be a general phenomenon.
A similar behavior of vacancies was previously found in MD simulations
of edge and screw dislocations in Al \citep{Pun09a}.

Although this study was focused on one particular high-angle GB chosen
as a model, the phenomena discussed are generic and should be relevant
to most GBs to one degree or another. Special low-energy GBs, such
as $\Sigma3$ coherent twin boundaries, may display specific features
such as weaker interaction with solutes and/or lower diffusivity.
The main conclusions of the work may not be applicable to them without
proper modifications.

\vspace{0.15in}

\textbf{Acknowledgement:} This work was supported by the National
Science Foundation, Division of Materials Research, under Award No.\,1708314.


\section*{{\LARGE{}Appendix}}

\begin{figure}[ht]
\noindent \centering{}\includegraphics[width=0.75\textwidth]{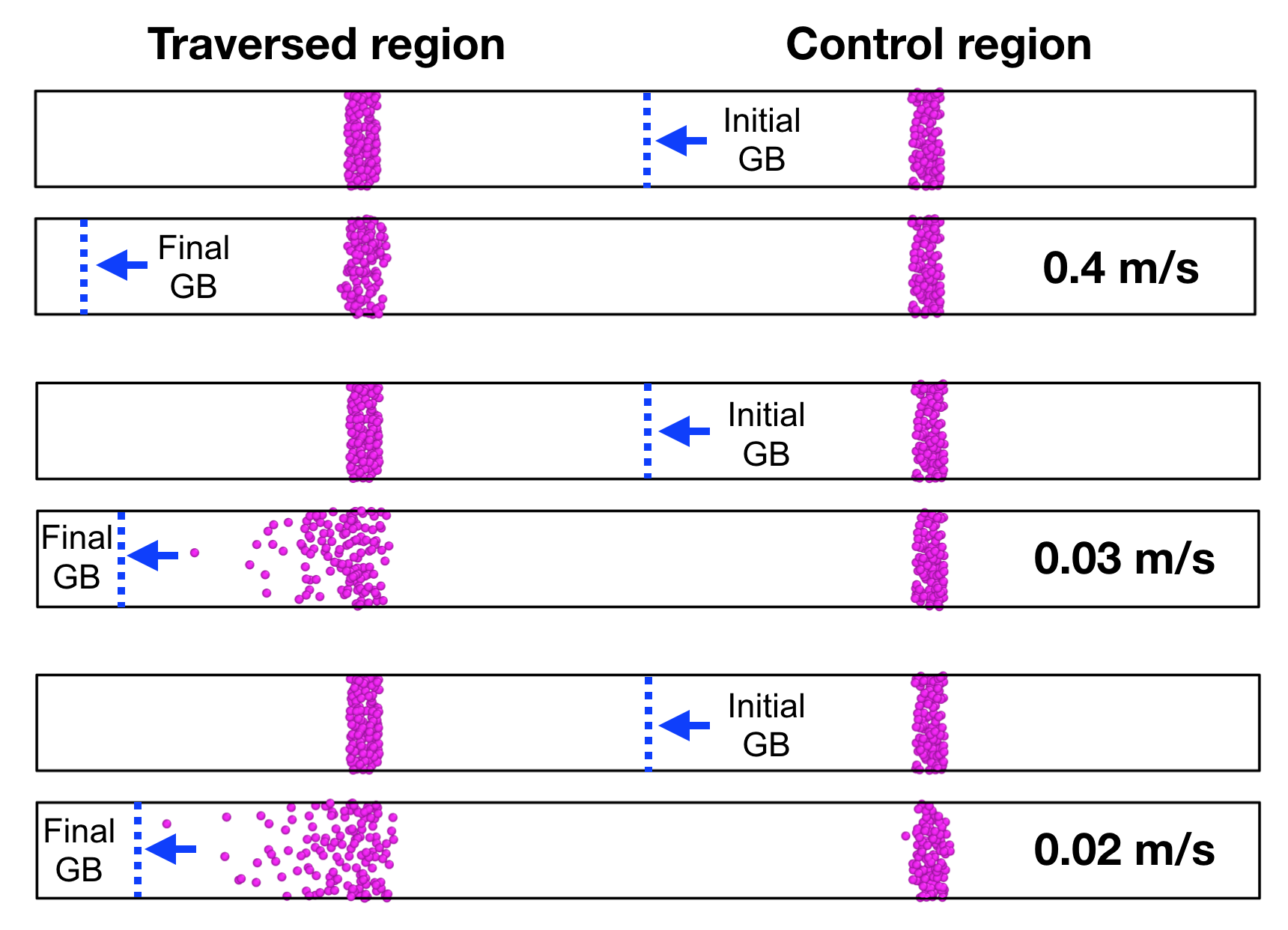}
\caption{Demonstration of the effect of GB motion on solute distribution. Ag
atoms within two 2 nm thick stripes are colored in magenta. The GB
moves to the left with the velocity indicated in the legend and crosses
the left-hand stripe, scattering the Ag atoms predominantly in the
direction of motion. The scattering asymmetry is stronger for the
smaller velocity. The right-hand stripe serves as a control region,
demonstrating the absence of lattice diffusion. The alloy composition
is Cu-2at.\%Ag.\label{fig:GBdrag}}
\end{figure}

\begin{figure}[ht]
\noindent \begin{centering}
(a) \includegraphics[width=0.6\textwidth]{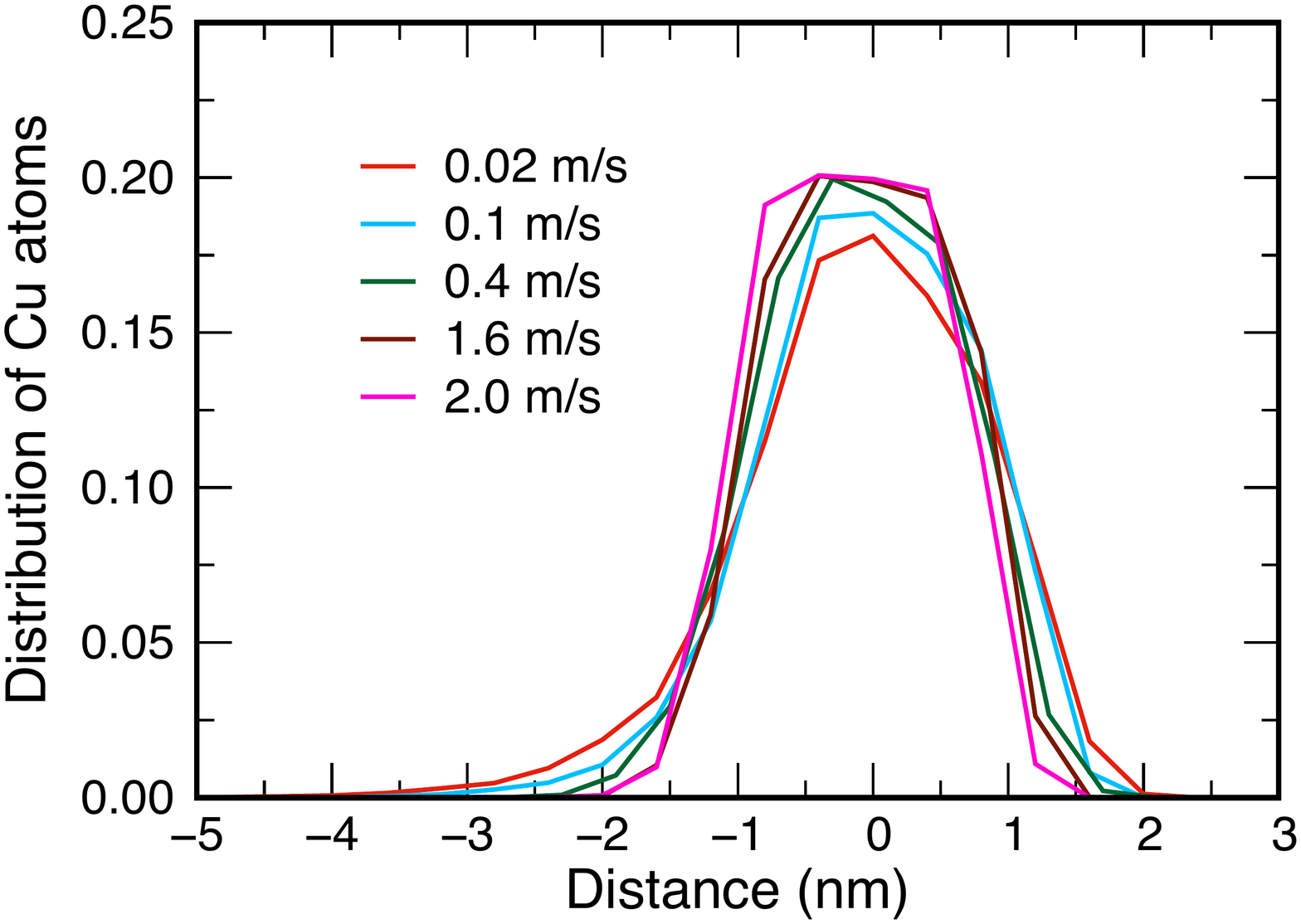} 
\par\end{centering}
\noindent \begin{centering}
\bigskip{}
\par\end{centering}
\noindent \centering{}(b) \includegraphics[width=0.6\textwidth]{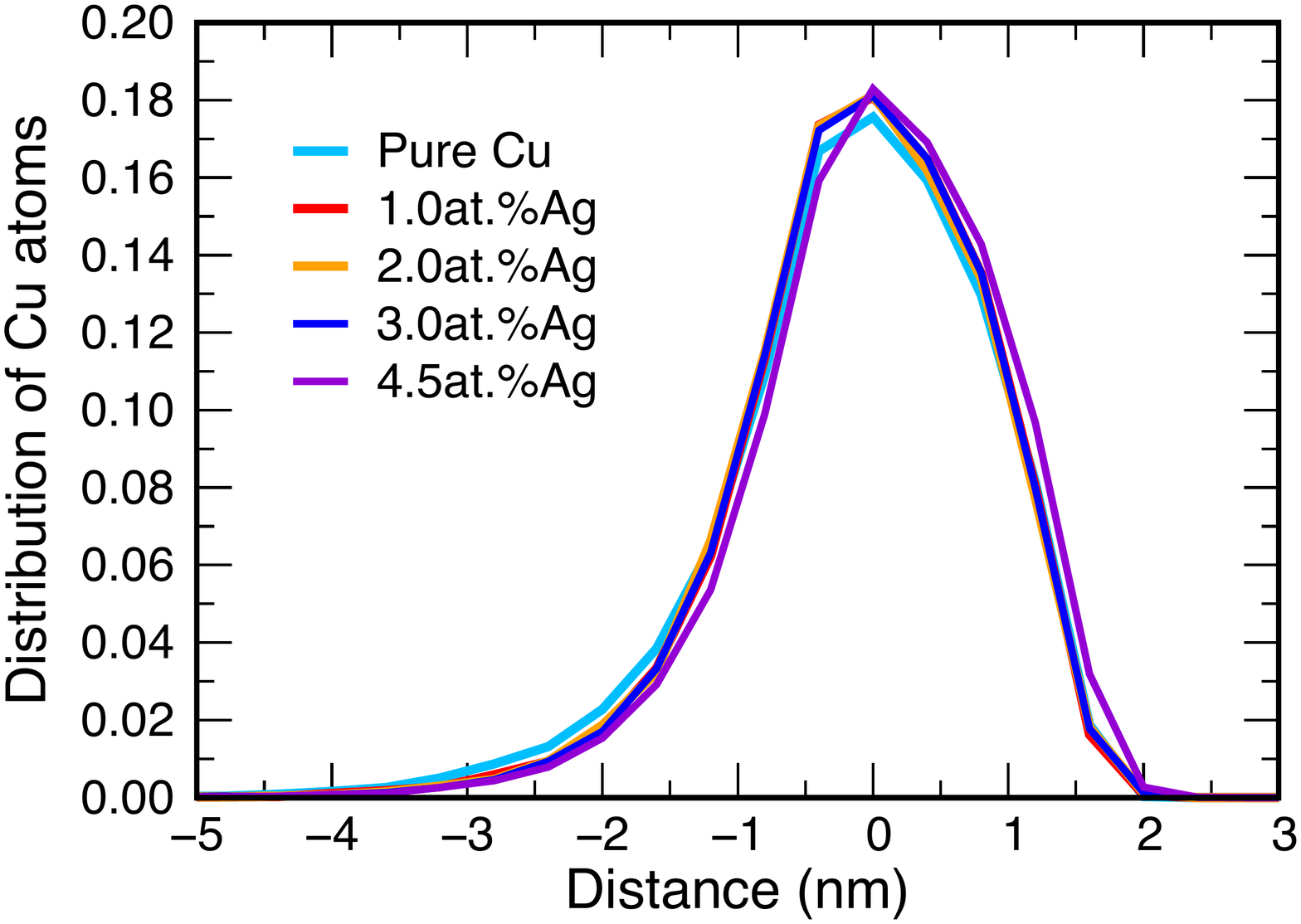}
\caption{Distribution profiles (normalized to unity) of Cu atoms in the final
state of a 2 nm stripe overrun by a moving GB: (a) for different GB
velocities at a fixed alloy composition (Cu-2at.\%Ag); (b) for different
alloy compositions at the fixed velocity of 0.02 m/s. The center of
the stripe is at $z=0$.\label{fig:GBdragB-1}}
\end{figure}

In this Appendix we provide additional evidence that the profile broadening
shown in Fig.~\ref{fig:GBdragB-1} is primarily caused by GB diffusion
with little or no effect of the solute drag. To this end, we numerically
solve the continuous diffusion equation 
\begin{equation}
\dfrac{\partial c}{\partial t}=D(z,t)\dfrac{\partial^{2}c}{\partial z^{2}}\label{eq:Diff-Eq}
\end{equation}
with the boundary condition $c(\pm\infty,t)=0$ and the initial condition
\begin{equation}
c(z,0)=\begin{cases}
1, & |z|\leq a,\\
0, & |z|>a.
\end{cases}\label{eq:stripe}
\end{equation}
Here, $c$ represents the concentration of Cu atoms selected within
a stripe of width $2a$ located at $z=0$ (cf.~Fig.~\ref{fig:GBdrag}).
These atoms can be thought of as isotope tracer atoms implanted into
the stripe. The diffusion coefficient is given by the moving Gaussian
\begin{equation}
D(z,t)=D_{0}\exp\left[-(z-z_{0}-vt)^{2}\right].\label{eq:Diff-Coeff}
\end{equation}
This function represents the enhanced diffusion coefficient in the
GB moving with the velocity $v$ starting from its initial position
at $z_{0}$. Thus, the diffusion coefficient tends to zero outside
the GB and reaches the largest value $D_{0}$ at the center of the
boundary. In the calculations shown here, the boundary moves to the
left starting from its initial position on the right of the stripe. 

\begin{figure}
\includegraphics[width=0.47\textwidth]{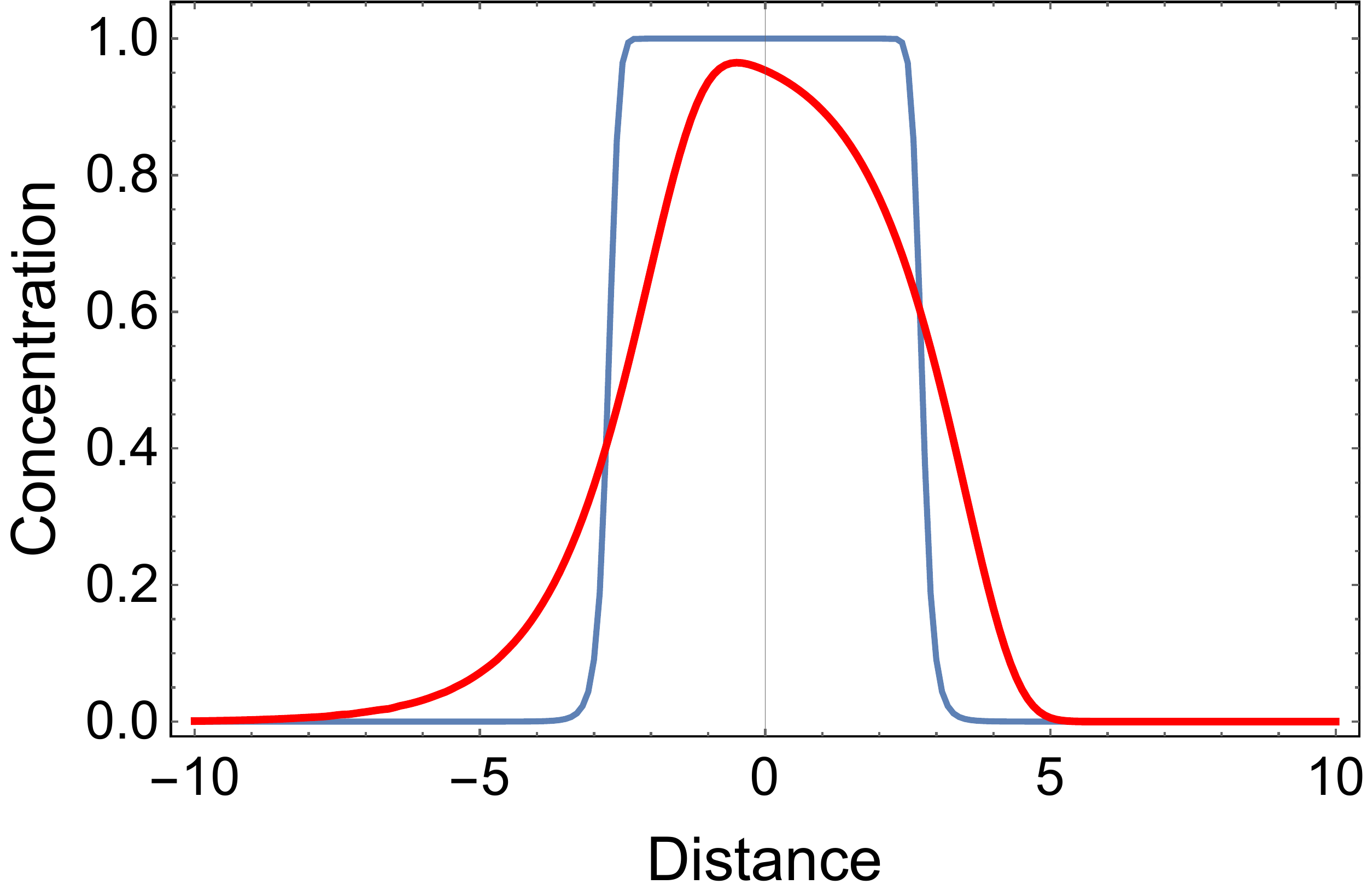}\quad{}\quad{}\includegraphics[width=0.47\textwidth]{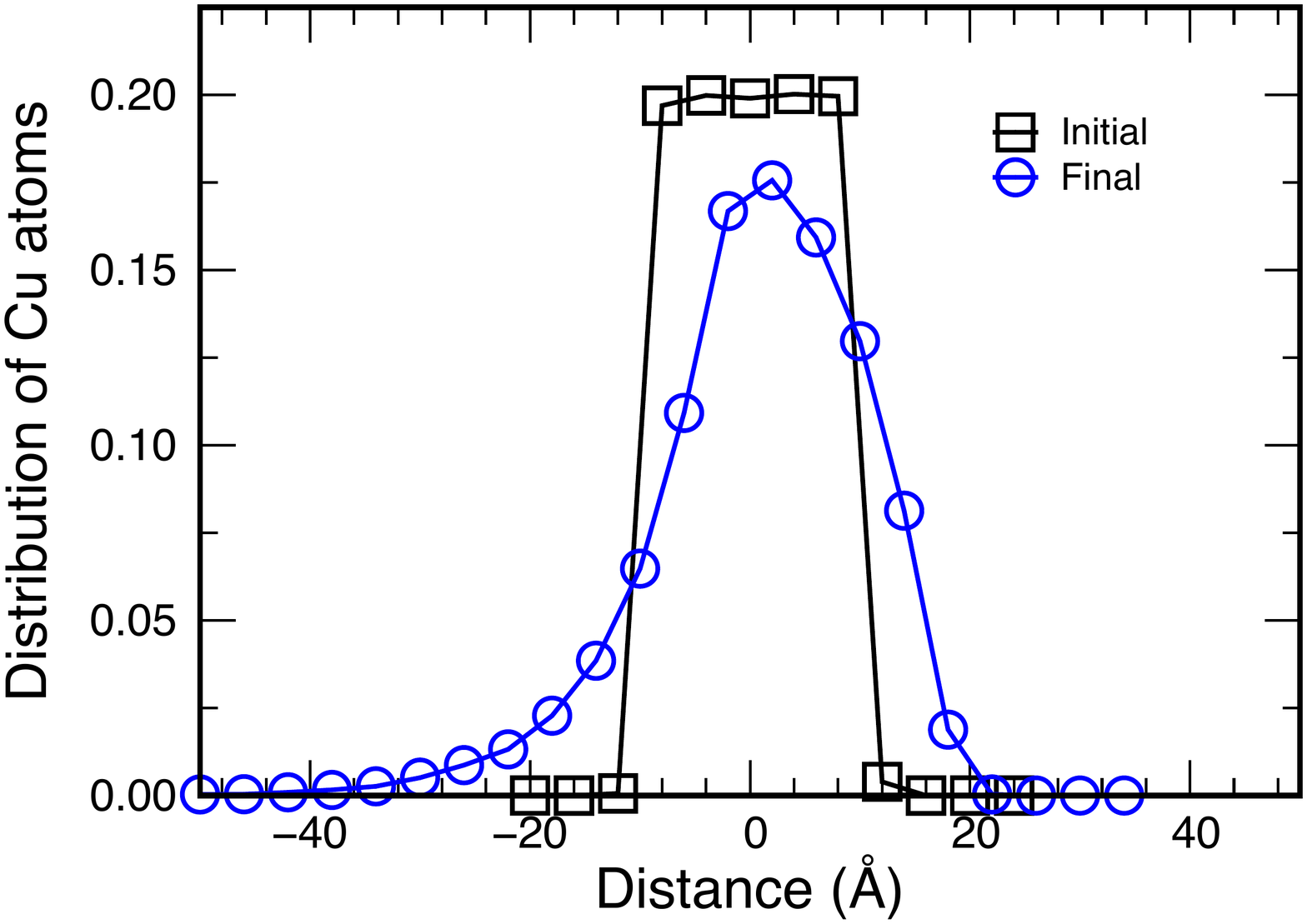}

\quad{}\quad{}(a)\quad{}\quad{}\quad{}\quad{}\quad{}\quad{}\quad{}\quad{}\quad{}\quad{}\quad{}\quad{}\quad{}\quad{}\quad{}\quad{}\quad{}\quad{}\quad{}\quad{}(b)

\bigskip{}

\bigskip{}

\includegraphics[width=0.47\textwidth]{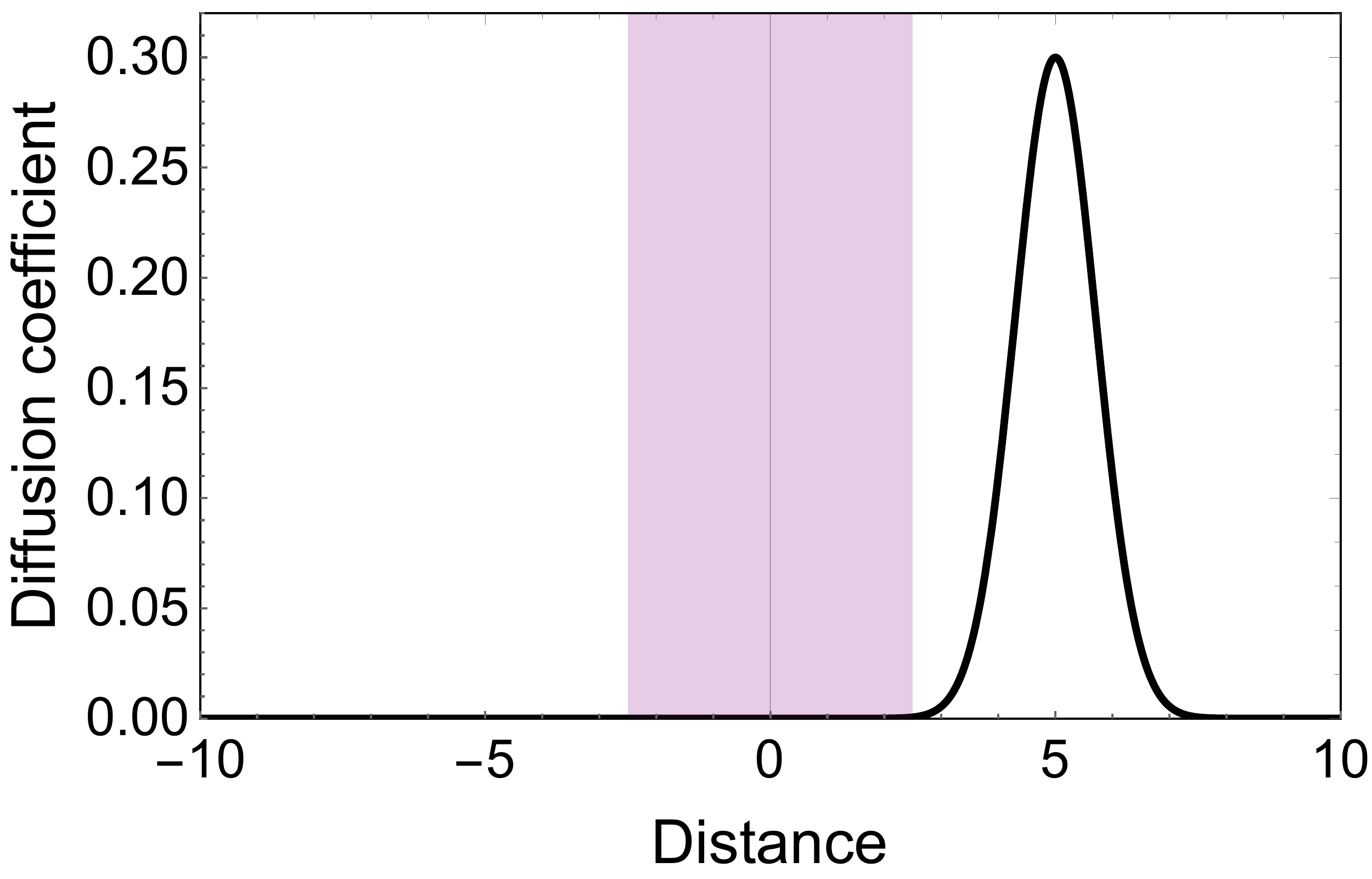}\quad{}\quad{}\includegraphics[width=0.47\textwidth]{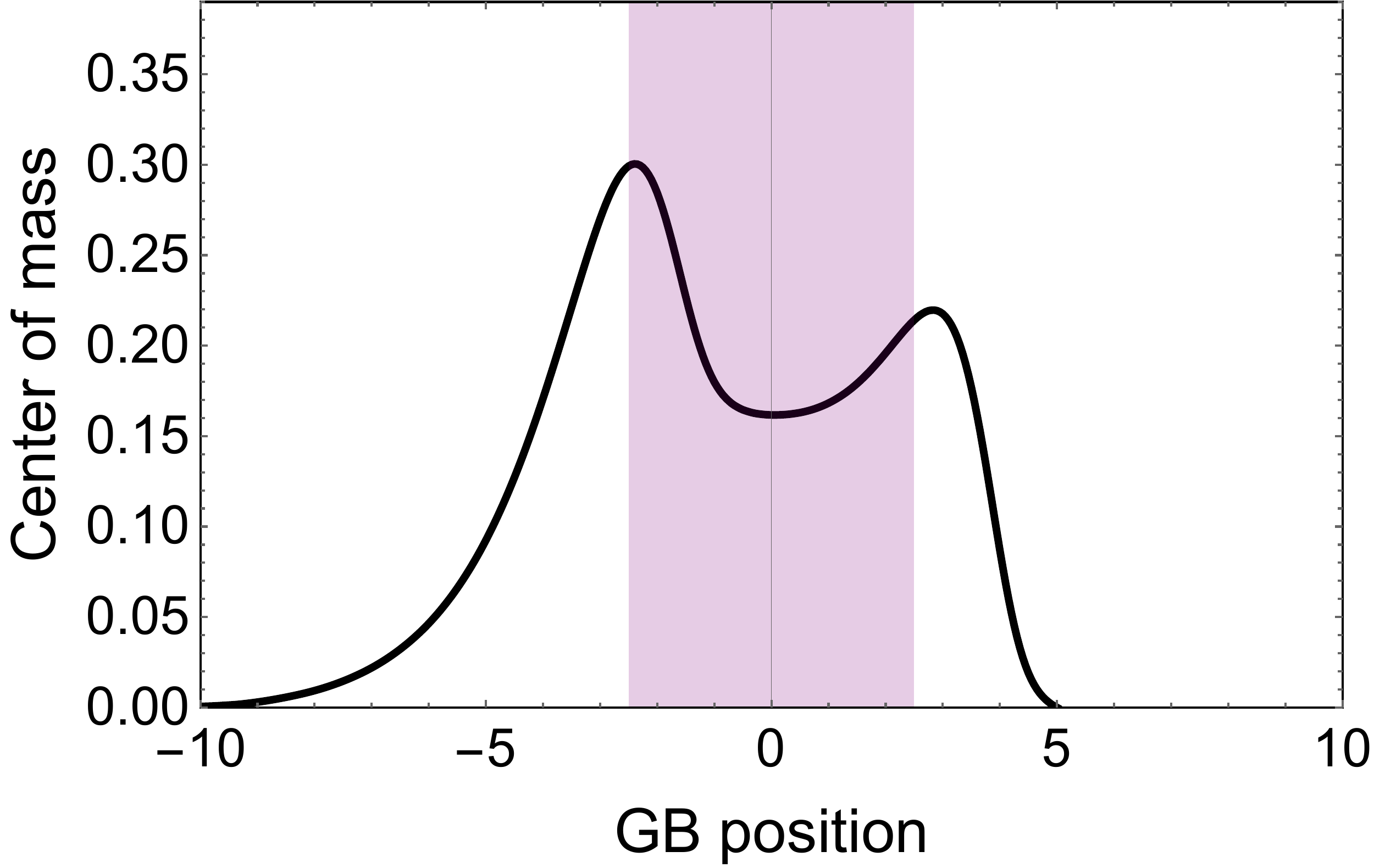}

\quad{}\quad{}(c)\quad{}\quad{}\quad{}\quad{}\quad{}\quad{}\quad{}\quad{}\quad{}\quad{}\quad{}\quad{}\quad{}\quad{}\quad{}\quad{}\quad{}\quad{}\quad{}\quad{}(d)

\caption{Concentration profiles before and after a moving GB passes through
a stripe of tracer atoms. (a) Predicted by the diffusion equation
(\ref{eq:Diff-Eq}) with the dimensionless parameters $a=2$, $z_{0}=5$,
$v=-0.3$ and $D_{0}=0.3$. The GB moves from right to left. The initial
profile is shown in blue, the final in red. (b) Similar profiles obtained
by atomistic simulations in pure Cu for the GB moving with the velocity
of 0.02 m/s. (c) The diffusion coefficient for the initial GB position
according to Eq.(\ref{eq:Diff-Coeff}). (d) Position of the center
of mass of the diffusing atoms as a function of GB position as the
latter traverses the stripe. In (c) and (d), the shaded area indicates
the stripe containing the diffusing atoms.\label{fig:Appendix}}
\end{figure}

Fig.~\ref{fig:Appendix}a compares the concentration profiles before
and after the boundary passes through the stripe, with the diffusion
coefficient shown in Fig.~\ref{fig:Appendix}c. The final shape of
the concentration profile is qualitatively similar to the profile
observed in the simulations (Fig.~\ref{fig:Appendix}b). Furthermore,
in Fig.~\ref{fig:Appendix}d we plot the position of the center of
mass of the diffusing atoms as a function of GB position. The center
of mass shows a small displacement toward the GB when the latter just
enters the stripe, slightly retreats while the GB is inside the stripe,
and finally moves slightly away from the boundary as the latter exits
the stripe. This produces the two peaks observed in the plot. However,
once the boundary passes through the stripe, the center of mass returns
to its original position. As expected physically, the diffusion process
does not produce a net displacement of the center of mass. Indeed,
the equations solved here only describe the enhancement of diffusion
by the moving boundary and do not include any interaction between
the diffusing atoms and the boundary. The temporary displacements
of the center of mass are smaller than the GB width and can be neglected
for practical intents and purposes.

\begin{figure}[ht]
\noindent \begin{centering}
(a) \includegraphics[width=0.6\textwidth]{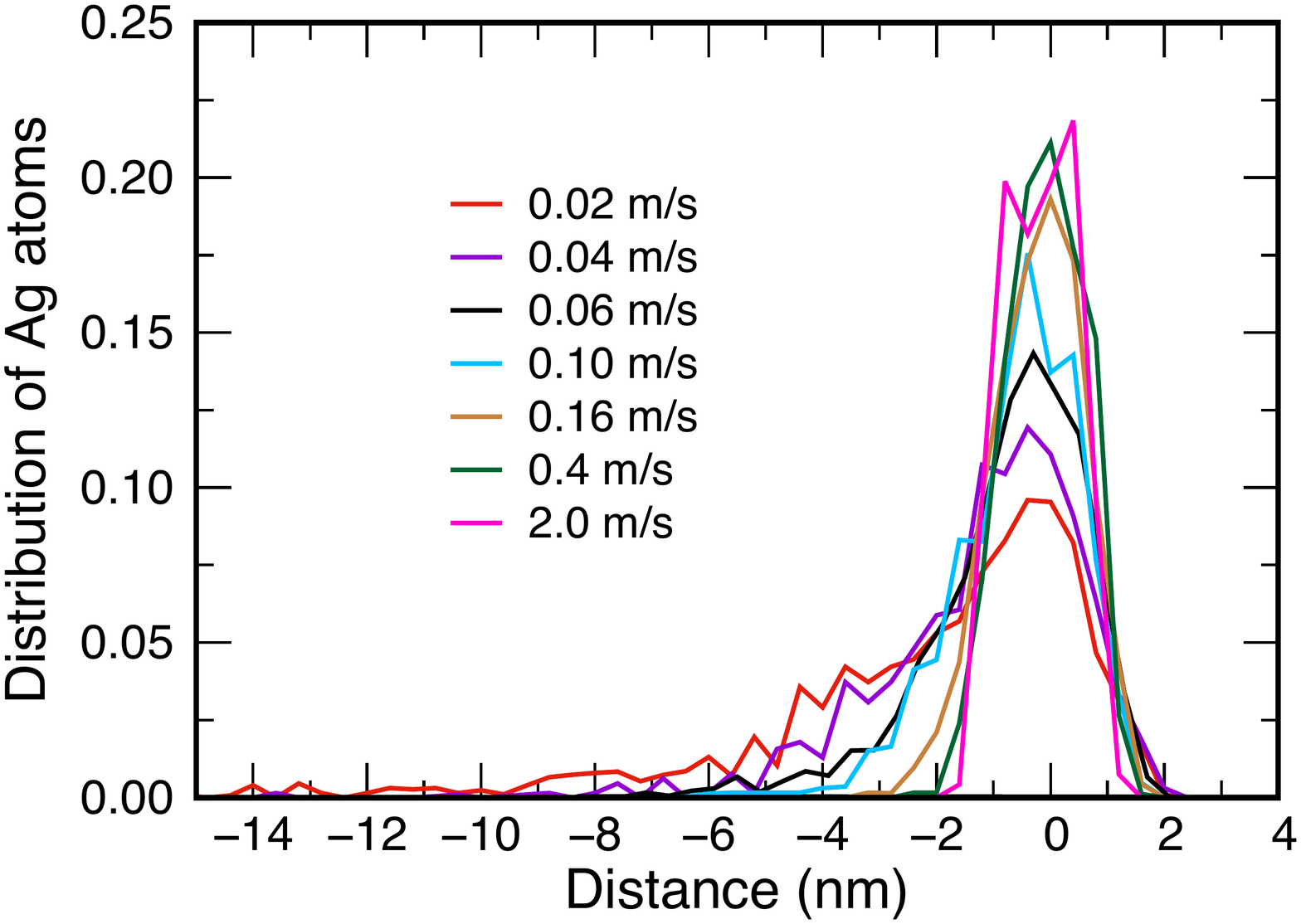}
\par\end{centering}
\noindent \begin{centering}
\bigskip{}
\par\end{centering}
\noindent \centering{}(b) \includegraphics[width=0.6\textwidth]{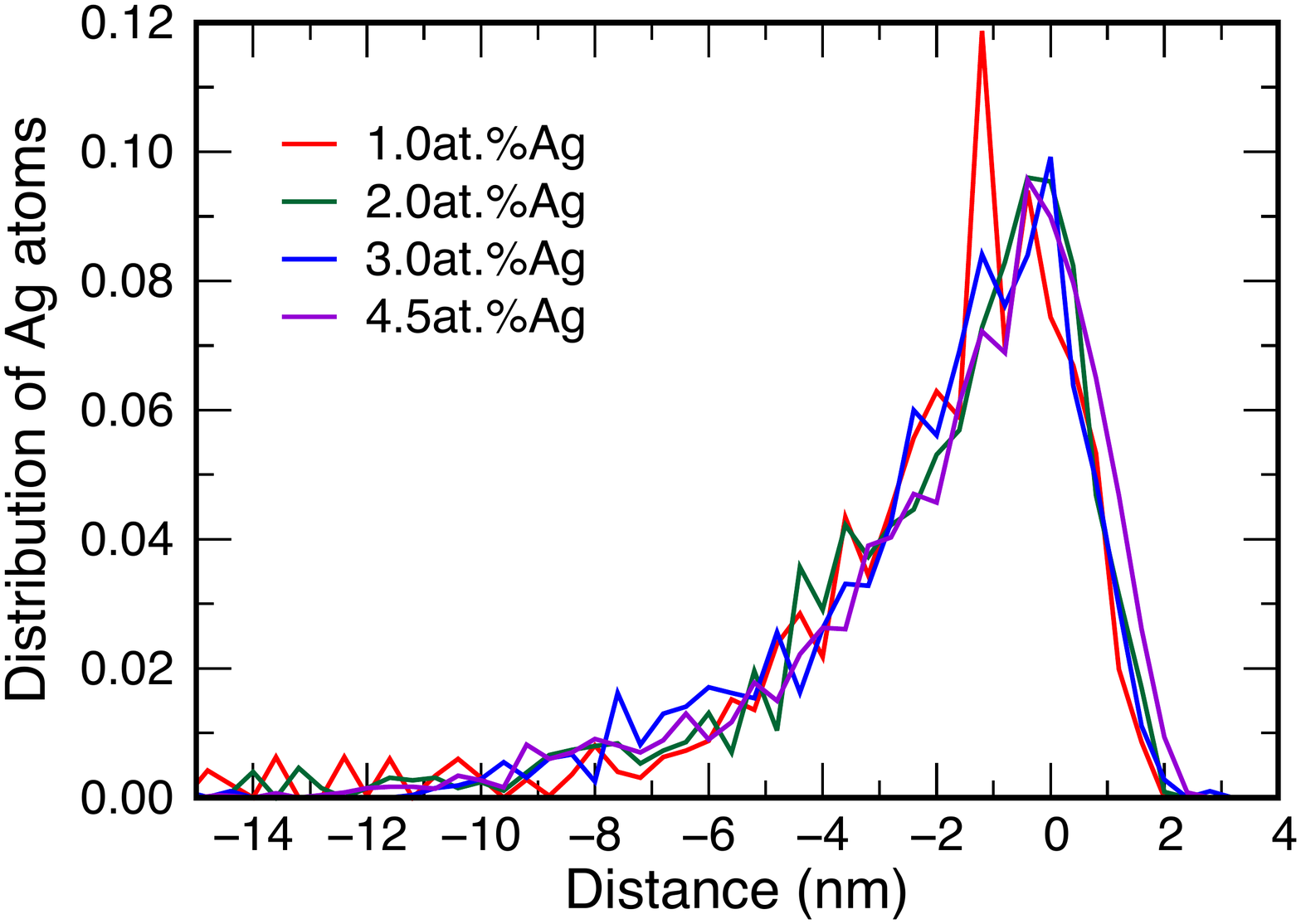}
\caption{Distribution profiles (normalized to unity) of Ag atoms in the final
state of a 2 nm stripe overrun by a moving GB: (a) for different GB
velocities at a fixed alloy composition (Cu-2at.\%Ag); (b) for different
alloy compositions at a fixed velocity of 0.02 m/s. The center of
the stripe is at $z=0$.\label{fig:GBdragB}}
\end{figure}

Thus, the concentration profiles of the Cu atoms observed in the simulations
(Fig.~\ref{fig:GBdragB-1}) are well-consistent with the purely diffusion
enhancement by the moving GB without any significant solute drag.
The drastic difference between these profiles and those for Ag atoms
(Fig.~\ref{fig:GBdragB}), featuring a long concentration tails,
confirms that the Ag atoms strongly interact with the moving GB and
are dragged by it in the direction of motion.

\newpage\clearpage{}

\end{document}